\documentclass[useAMS,usenatbib]{mn2e}

\usepackage[]{natbib}
\usepackage{graphicx}
\usepackage{multirow}

\title[CVSO\,30]{YETI observations of the young transiting planet candidate CVSO\,30\,b}
\author[St. Raetz et al.]{St. Raetz$^{1}$\thanks{E-mail:sraetz@cosmos.esa.int},  T.O.B. Schmidt$^{2}$, S. Czesla$^{2}$, T.Klocov\'a$^{2}$, L. Holmes$^{3}$, R. Errmann$^{4,5}$, \newauthor M. Kitze$^{4,6}$, M. Fern\'{a}ndez$^{7}$, A. Sota$^{7}$, C. Brice\~{n}o$^{8}$, J. Hern\'andez$^{9}$,  J. J. Downes$^{9}$, \newauthor D.P. Dimitrov$^{10}$, D. Kjurkchieva$^{11}$, V. Radeva$^{11}$, Z.-Y. Wu$^{12}$ , X. Zhou$^{12}$, H. Takahashi$^{13}$, \newauthor T. Henych$^{14}$,  M. Seeliger$^{4}$, M. Mugrauer$^{4}$, Ch. Adam$^{4}$, C. Marka$^{15}$, J.G. Schmidt$^{4}$, \newauthor M.M. Hohle$^{16}$,  Ch. Ginski$^{17}$, , T. Pribulla$^{18}$, L. Trepl$^{4}$, M. Moualla$^{19}$, N. Pawellek$^{4}$, \newauthor J. Gelszinnis$^{20}$, S. Buder$^{4}$, S. Masda$^{4}$, G. Maciejewski$^{21}$ and R. Neuh\"{a}user$^{4}$ \\
$^{1}$Scientific Support Office, Directorate of Science, European Space Research and Technology Centre (ESA/ESTEC),\\ Keplerlaan 1, 2201 AZ Noordwijk, The Netherlands\\
$^{2}$Hamburger Sternwarte, Gojenbergsweg 112, D-21029 Hamburg, Germany\\
$^{3}$Department of Physics and Astronomy, University of Leicester, University Road, Leicester, LE1 7RH, UK \\
$^{4}$Astrophysikalisches Institut und Universit\"{a}ts-Sternwarte, Schillerg\"{a}\ss{}chen 2-3, D-07745 Jena, Germany\\
$^{5}$Abbe Center of Photonics, Friedrich-Schiller-Universit\"{a}t Jena, Max-Wien-Platz 1, 07743 Jena, Germany \\
$^{6}$Universit\"{a}t Rostock, Institut für Physik, D-18051 Rostock, Germany\\
$^{7}$Instituto de Astrof\'{\i}sica de Andaluc\'{\i}a, CSIC, Apdo. 3004, 18080 Granada, Spain\\
$^{8}$Cerro Tololo Inter-American Observatory CTIO/AURA/NOAO, Colina El Pino s/n. Casilla 603, 1700000 La Serena, Chile \\
$^{9}$Centro de Investigaciones de Astronom\'{\i}a (CIDA), Apdo. Postal 264, M\'{e}rida 5101-A, Venezuela \\
$^{10}$Institute of Astronomy and NAO, Bulg. Acad. Sc., 72 Tsarigradsko Chaussee Blvd., 1784 Sofia, Bulgarian \\
$^{11}$Department of Physics, Shumen University, 9700 Shumen, Bulgaria.\\
$^{12}$Key Laboratory of Optical Astronomy, NAO, Chinese Academy of Sciences, 20A Datun Road, Beijing 100012, China\\
$^{13}$Gunma Astronomical Observatory, 6860-86 Nakayama, Takayama-mura, Agatsuma-gun, Gunma 377-0702 Japan \\
$^{14}$Astronomical Institute, Academy of Sciences of the Czech Republic, Fri\v{c}ova 298, CZ-25165 Ond\v{r}ejov, Czech Republic\\
$^{15}$Instituto Radioastronom\'ia Milim\'{e}trica (IRAM), Avenida Divina Pastora 7, 18012 Granada, Spain \\
$^{16}$Gene Center of the Ludwig-Maximilians-University, Feodor-Lynen-Strasse 25, 81377 Munich, Germany \\
$^{17}$Sterrewacht Leiden, PO Box 9513, Niels Bohrweg 2, NL-2300RA Leiden, the Netherlands \\
$^{18}$Astronomical Institute, Slovak Academy of Sciences, 059 60 Tatransk\'{a} Lomnica, Slovakia \\
$^{19}$Department of physics, Faculty of science, Tishreen University, Lattakia, Syria \\
$^{20}$Th\"{u}ringer Landessternwarte Tautenburg, Sternwarte 5, D-07778 Tautenburg, Germany \\
$^{21}$Centre for Astronomy, Faculty of Physics, Astronomy and Informatics, Nicolaus Copernicus University, Grudziadzka 5, 87-100 Torun, \\Poland
 }

\begin{document}

\date{Accepted 2016 May 11. Received 2016 May 11; in original form 2015 September 18}

\pagerange{\pageref{firstpage}--\pageref{lastpage}} \pubyear{2002}

\maketitle

\label{firstpage}

\begin{abstract}
CVSO\,30 is a unique young low-mass system, because, for the first time, a close-in transiting and a wide directly imaged planet candidates are found around a common host star. The inner companion, CVSO\,30\,b, is the first possible young transiting planet orbiting a previously known weak-lined T-Tauri star. With five telescopes of the 'Young Exoplanet Transit Initiative' (YETI) located in Asia, Europe and South America we monitored CVSO\,30 over three years in a total of 144 nights and detected 33 fading events. In two more seasons we carried out follow-up observations with three telescopes. We can confirm that there is a change in the shape of the fading event between different observations and that the fading event even disappears and reappears. A total of 38 fading event light curves were simultaneously modelled. We derived the planetary, stellar, and geometrical properties of the system and found them slightly smaller but in agreement with the values from the discovery paper. The period of the fading event was found to be 1.36\,s shorter and 100 times more precise than the previous published value. If CVSO\,30\,b would be a giant planet on a precessing orbit, which we cannot confirm, yet, the precession period may be shorter than previously thought. But if confirmed as a planet it would be the youngest transiting planet ever detected and will provide important constraints on planet formation and migration time-scales.

\end{abstract}


\section{Introduction}

During the last two decades the existence of other planetary systems has gone from speculation to fact.  With well over a thousand planets discovered so far, one of the key questions is how planets are formed. The two main scenarios currently proposed are a stellar-like formation in the protostellar cloud through gravitational collapse \citep[e.g.][]{2001ApJ...551L.167B} or the formation in the circumstellar disc \citep[e.g.][]{1983RvGSP..21..206W}. For the latter scenario two models have been proposed, the core-accretion scenario \citep{1969Icar...10..109S,1973ApJ...183.1051G,1996Icar..124...62P} and the disc instability scenario \citep{1978M&P....18....5C,1997Sci...276.1836B}. Since the discovery of the first very close-in giant exoplanet around a main-sequence star \citep{1995Natur.378..355M} it was argued that those planets cannot have formed in-situ but have formed further outwards and than moved inwards by planet-disc migration. An open problem in planet formation  are the time-scales. According to the core-accretion scenario, a core is built by collisions of planetesimals, gas from the surrounding disc is accreted and a gas giant is formed. However, in this scenario the time to form a gas giant is close to the gas depletion timescale of the discs \citep{2001ApJ...553L.153H}, while, according to the disc instability scenario, gas giants can be quickly formed by disc fragmentation before the gas in the disc is depleted \citep{2007lyot.confE..18M}. Recently, it was reported that the core-accretion scenario can overcome the time-scale problem by planetesimal formation via 'pebble' concentration and gravitational collapse. This so-called 'pebble accretion' model can produce cores of ten Earth masses in a few thousand years \citep{2015Natur.524..322L}. \\ As of today (2016 April 14) 2107 extrasolar planets (candidates) in 1349 planetary system are listed in the ``The Extrasolar Planets Encyclopaedia'' (exoplanet.eu). Almost all the planets (and host stars) are, however, Gyr old, making it difficult to study planet formation. Planets (candidates) around pre-main sequence (PMS) stars have been discovered so far only with the direct imaging technique. However, because these planets are usually on very wide orbits around their host star it is not possible to determine their mass dynamically. Hence, their planetary status is model-dependent and still uncertain. Young transiting planets are of great importance for the study of planet formation since their observed light curves (LCs) directly yield planetary, stellar and geometrical properties. Therefore it is possible to test evolutionary models and, hence, to distinguish between planet formation scenarios. \\To constrain the limits for the time-scales of planet formation and migration we established the ``Young Exoplanet Transit Initiative'' (YETI), a search for transiting planets in young open clusters. The motivation, observing strategy, target cluster selection, and first results of our first target cluster Trumpler\,37 can be found in \citet{2011AN....332..547N} and \citet{2013AN....334..673E,2014AN....335..345E}. In summary, YETI is a network of small to medium size telescopes (0.2 to 2.6\,m) spread worldwide at different longitudes. The telescope network enables the observation of the targets continuously for several days in order not to miss any transit.
\begin{table}
\caption{Physical and orbital properties of the CVSO\,30 system summarized from literature.}
\label{Werte_CVSO30}
\begin{tabular}{ccc}
\hline \hline
Parameter & Value & Ref \\ \hline
\multicolumn{3}{c}{stellar parameters} \\ \hline 
Mass star $M_{\mathrm{A}}$ (Baraffe) [M$_{\odot}$] & 0.44 & [1] \\
Mass star $M_{\mathrm{A}}$ (Siess) [M$_{\odot}$] & 0.34 & [1] \\
Radius star $R_{\mathrm{A}}$  [R$_{\odot}$] & 1.39 & [1] \\
Effective temperature $T_{\mathrm{eff}}$  [K] & 3470 & [1] \\
Distance $d$ [pc] & 323$^{+233}_{-96}$ & [1] \\ 
Age [Myr] & 2.39$^{+3.41}_{-2.05}$ & [5] \\
v\,sin($i$) [km\,s$^{-1}$] & 80.6$\pm$8.1 & [2]\\
Spectral type &  M3 & [1] \\
class &  WTTS & [1]\\
$V$ [mag] & 16.26 & [1] \\
$R$ [mag] & 15.19 & [2] \\
$I$ [mag] & 13.74 & [2] \\
$J$ [mag] & 12.232$\pm$0.028 & [4] \\
$H$ [mag] & 11.559$\pm$0.026 & [4] \\
$K_{S}$ [mag] & 11.357$\pm$0.021 & [4] \\\hline
\multicolumn{3}{c}{planetary parameters CVSO\,30\,b} \\ \hline 
Epoch zero transit time $T_{0}$  [d] & 2455543.9402 &  \\
 &  \hspace{0.7cm}$\pm$0.0008 &  [2] \\
Orbital period $P_{\mathrm{b}}$  [d] & 0.448413$\pm$0.000040 & [2] \\
Semi-major axis $a_{\mathrm{b}}$ [au] & 0.00838$\pm$0.00072 & [2] \\
Inclination $i$  [$^{\circ}$] & 61.8$\pm$3.7 & [2] \\
Radius planet $R_{\mathrm{b}}$  [$R_{\mathrm{Jup}}$] & 1.91$\pm$0.21 & [2] \\
 & 1.64/1.68$\pm$0.07$^{a}$ & [3] \\
Mass planet $M_{\mathrm{b}}$  [$M_{\mathrm{Jup}}$] & $<$5.5$\pm$1.4 & [2] \\
 & 3.0$\pm$0.2/3.6$\pm$0.38$^{a}$  & [3] \\
spin-orbit angle $\varphi$  [$^{\circ}$] & 69$\pm$2/73.1$\pm$0.6$^{a}$ & [3] \\\hline
\multicolumn{3}{c}{planetary parameters CVSO\,30\,c} \\ \hline 
Orbital period $P_{\mathrm{c}}$  [yr] & $\sim$27250 & [5] \\
Semi-major axis $a_{\mathrm{c}}$ [au] & 662$\pm$96 & [5] \\
Effective temperature $T_{\mathrm{eff}}$  [K] & 1600$^{+120}_{-300}$ & [5] \\
log\,$g_{\mathrm{c}}$ & 3.6$^{+1.4}_{-0.6}$ & [5] \\
Radius planet $R_{\mathrm{c}}$  [$R_{\mathrm{Jup}}$] & 1.63$^{+0.87}_{-0.34}$ & [5] \\
Mass planet $M_{\mathrm{c}}$  [$M_{\mathrm{Jup}}$] & $4.7^{+3.6}_{-2.0}$ & [5] \\
J band (differential$^{b}$) [mag] & 7.385$\pm$0.045 & [5] \\
H band (differential$^{b}$) [mag] & 7.243$\pm$0.014 & [5] \\
K$_{S}$ band (differential$^{b}$) [mag] & 7.351$\pm$0.022 & [5] \\
\hline \hline
\end{tabular}
\\$^{a}$different values due to using the stellar mass derived either with Baraffe or Siess models
\\$^{b}$Difference between host star and CVSO\,30\,c
\\ References: [1] \citet{2005AJ....129..907B}, [2] \citet{2012ApJ...755...42V}, and [3] \citet{2013ApJ...774...53B} [4] \citet{2006AJ....131.1163S} [5] \citet{Schmidt}
\end{table}
\\ Young open clusters provide an ideal environment for the search for young extrasolar planets and to study stellar variability, since they feature a relatively large number of stars of the same known age and metallicity at the same distance.\\ One target of YETI is the young open cluster 25\,Ori in the nearby Orion OB1 association. It was discovered by \citet{2007ApJ...661.1119B} and contains $>$\,200 low-mass PMS stars concentrated within $\sim$\,1$^{\circ}$ around the early B-type star 25\,Ori. The Hipparcos stars in the cluster yielded a distance to 25\,Ori of $\sim$\,330\,pc. The position of the low mass members in the colour-magnitude diagram corresponds to an isochronal age of $\sim$7-10 Myrs. 25\,Ori is the most populated cluster in this age range known within 500\,pc and, hence, is an excellent laboratory to study the early evolution of sun-like stars, protoplanetary discs, and planet formation. Our observations of 25\,Ori started in January 2010. During the (northern) winter 2010/2011 25\,Ori became a target of YETI where it was monitored for three consecutive years with up to 13 telescopes located in Europe, Asia and America. As a result we confirm the presence of transit-like flux drops first reported by \citet{2012ApJ...755...42V}. Here we present our YETI (four telescopes) and photometric follow-up observations of CVSO\,30.

\section{CVSO\,30 in 25\,Ori}

\begin{table*}
\caption{Observatories and instruments which monitored (first five lines) or followed-up CVSO\,30.}
\label{CCD_Kameras}
\begin{tabular}{ccccccccc}
\hline \hline
Observatory & Long. (E) & Lat. (N) & Altitude & Mirror $\diameter$ & Camera & \# Pixel & Pixel scale & FoV\\ 
& [$^{\circ}$] & [$^{\circ}$] & [m] & [m] & & &  [$''/pix$] & [$'$]\\ \hline
Gunma/Japan & 139.0 & 36.6 & 885 & 1.50 & Andor DW 432 & 1250\,x\,1152 & 0.57 &  12.5\,x\,12.5 \\
Xinglong/China & 117.6 & 40.4 & 960 & 0.90$^{b}$ & E2V CCD203-82 & 4096\,x\,4096 & 1.38 & 94.0\,x\,94.0 \\
Rozhen/Bulgaria & 24.7 & 41.7 & 1759 & 0.70$^{a}$ & FLI\,ProLine\,16803 & 4096\,x\,4096 & 1.08 & 73.8\,x\,73.8 \\
Jena/Germany  & 11.5 & 50.9 & 370 & 0.90$^{b}$ & E2V CCD42-10  & 2048\,x\,2048 & 1.55 & 52.8\,x\,52.8 \\
  &  &  &  &  & (STK)$^{c}$  &  &  &  \\
CIDA/Venezuela & 289.1 & 8.8 & 3600 & 1.00 & FLI\,ProLine\,4240 & 2048\,x\,2048 & 0.54 & 19.2\,x\,19.2 \\ \hline
Sierra Nevada/Spain & 356.6 & 30.1 & 2896 & 1.50 & VersArray:2048B & 2048\,x\,2048 & 0.23 & 7.8\,x\,7.8  \\
 & & & & 0.90 & VersArray:2048B &  2048\,x\,2048 & 0.39 & 13.2\,x\,13.2 \\
La Silla/Chile & 289.3 & -29.3 & 2335 & 2.2 & Wide Field Imager &  8 times  &  &  \\
  &  &  &  &  & (WFI)  & 2142\,x\,4128 & 0.24 & 34.0\,x\,32.7 \\
\hline \hline
\end{tabular}
\\
$^{a}$0.50\,m in Schmidt mode, $^{b}$0.60\,m in Schmidt mode, $^{c}$ \citet{2010AN....331..449M}
\end{table*}

CVSO 30 (2MASS J05250755+0134243, PTFO 8-8695) was first identified as a weak-line T-Tauri star in the large-scale, multiepoch CIDA Variability Survey of Orion OB1 \citep{2005AJ....129..907B}. The star of spectral type M3 is located in the OB1a sub-association at an average distance of 357$\pm$52\,pc \citep{2014MNRAS.444.1793D} and is a member of 25\,Ori \citep{2007ApJ...661.1119B}. The fast rotating PMS star CVSO\,30 with an effective temperature of $\sim$\,3470\,K is one of the youngest members of 25\,Ori. Isochrone fitting yielded an age of $\sim$2.4\,Myr and a mass of 0.34 to 0.44\,M$_{\odot}$ (depending on the used stellar evolutionary model). The LC of the young star CVSO\,30 is dominated by stellar variability as expected for a PMS object. Within our data set we find the amplitude of light variation for the R\,=\,15.2\,mag star varying up to $\sim$\,0.1 mag (excluding occasional flares).\\ \citet{2012ApJ...755...42V} first discovered the fading events of CVSO\,30 in the data of the Palomar Transient Factory (PTF) Orion project. The survey used the Palomar 48$''$ Samuel Oschin telescope to monitor a 7.26\,$\mathrm{deg^{2}}$ region centred around the young open cluster 25\,Ori \citep{2011AJ....142...60V}. The field was observed for 14 nights between 2009 December 1 and 2010 January 15 and another seven nights in 2010 December. \\ The transiting planet candidate CVS0\,30 shows a typical transit-like LC with an period close to or synchronous with the stellar rotation period. Every $\sim$\,0.4484\,d the brightness drops for $\sim$\,100\,min by $\sim$\,37\,mmag. With one of the shortest periods known so far and the very small orbital radius of around twice the stellar radius it appears to be at or within the stellar Roche limiting radius. Therefore CVSO\,30\,b could be subject to mass loss or disintegration due to tidal forces induced by its host star. \\ An interesting feature of the LC of CVSO 30 was mentioned by \citet{2012ApJ...755...42V}. In their two sets of LCs (2009 and 2010) it can clearly be seen that there is an overall change in the shape of the fading event between the two years. \citet{2013ApJ...774...53B} showed that the unusual LC shapes of CVSO\,30 and their variation in \citet{2012ApJ...755...42V} can be explained by a precessing planet transiting a gravity-darkened star \footnote{if a star rotates fast enough to become oblate it shows a higher surface gravity on the poles than on the equator, and thus a higher temperature and brightness \citep{1924MNRAS..84..665V}}. From their modelling they derived a precession period of 300 to 600\,d assuming the spin-orbit to be synchronously locked. As a consequence of the precession the fading event is expected to disappear for a period of time. \citet{2015arXiv150604829K} reanalysed the LCs along with their own observations at the Koyama Astronomical Observatory. Their precession modelling, without requiring the spin-orbit synchronous condition, resulted in three possible precession periods (827,475, and 199\,d), the latter one is preferred by their observations. \citet{2016MNRAS.457.3769H} repeated the precession modelling using improved treatments of stellar geometry, surface intensities, and gravity darkening. They found that the LCs can be reproduced but their solution requires a near-critical stellar rotation and a significant photometric variability which disagrees the observations. Therefore they claimed that 'an exoplanet transiting a precessing, gravity-darkened star' may not be the correct explanation.  \\  \citet{2012ApJ...755...42V} obtained radial velocity (RV) measurements in the visual wavelength range using the HRS on the Hobby Eberly Telescope and HIRES on the Keck I Telescope. Their fitted transit model (orbital elements were fixed to the photometric derived values) appears significantly out of phase with the data. Their best-fitting model shows a phase-offset from the transit ephemeris. They conclude that the RV signal most likely arises because of spot effects modulated by the stellar rotation, where the amplitude of the spot effect is at least comparable or even greater than the signal from the planet. As a result they estimate an upper mass limit for the planet of $M_{Pl}\leq5.5\pm1.4M_{Jup}$. Although \citet{2012ApJ...755...42V} could not definitively rule out potential false positives \citet{2013ApJ...774...53B} could not find any false-positive scenario that could reproduce the combination of gravity darkening and nodal precession that is seen in the system. From these investigations \citet{2013ApJ...774...53B} could narrow down the mass range of CVSO\,30\,b to 3.0 to 3.6 $M_{Jup}$. \\ Photometric follow-up observations of \citet{2015arXiv150608719C} with the Las Cumbres Observatory Global Telescope Network (LCOGT) and Spitzer show the expected dis- and reappearance of the fading event, while the RV follow-up observations with Keck NIRSPEC failed to detect the Rossiter-McLaughlin effect as well as the planetary signal. From a comparison with the precession models they concluded that model and observations are not in perfect agreement and that the data are currently insufficient to confirm the planetary status of CVSO\,30\,b.\\ Multicolour photometry of CVSO\,30 from \citet{2015MNRAS.450.3991K} in six nights in January 2015 showed no signs of a fading event. He concluded that the dips in the LC could be either part of a complicated non-sinusoidal variability or a temporary absence of the fading events due to precession of the orbit as previously claimed. \\ \citet{2015arXiv150902176Y} presented three tests of the planet hypothesis. They observed 26 different fading events with the 1.2\,m telescope at the Fred Lawrence Whipple Observatory (FLWO) and the 0.6\,m TRAnsiting Planets and PlanetesImals Small Telescope (TRAPPIST) between 2012 and 2015, some of them simultaneously in different filters. They also carried out ground-based infrared observations with one of the Magellan 6.5\,m telescopes and re-analysed the Spitzer data, already reported in \citet{2015arXiv150608719C}, in order to identify the secondary eclipse. Furthermore, to detect the Rossiter-McLaughlin effect they obtained high-resolution spectroscopy with HIRES. They created five hypotheses to explain the existence of the brightness dip but disfavour the giant-planet model because all three tests failed to confirm the planetary nature of CVSO\,30\,b. \\ Recently, \citet{Schmidt} reported on the direct detection of a wide separation ($\sim$660\,au) planet candidate around CVSO\,30. Hence, CVSO\,30 is the first system harbouring both a close-in transiting and wide separation direct imaging planet candidates. From spectroscopic observations \citet{Schmidt} deduced a mass for CVSO\,30\,c of $4.7^{+3.6}_{-2.0}M_{Jup}$ which is very close to the mass of the putative planet candidate CVSO\,30\,b. The properties of the system including the host star and the two planet candidates are summarized in Table~\ref{Werte_CVSO30}. This system will give us the matchless opportunity to study planet formation and migration theories such as the planet-planet scattering that may be responsible for massive close-in planets \citep{1996Natur.384..619W}.

\section{Observation and Data Reduction}
\label{data_reduction}

\begin{table}
\centering
\caption{Summary of the CVSO\,30 monitoring observations in the period from 2010 January to 2013 February.}
\label{Beobachtungen}
\begin{tabular}{clcc}
\hline \hline
Observatory & Run* & Date & nights \\ \hline\hline
Jena/Germany & S00 & outside campaign & 10 \\
& S01-1 & 2010 Dec 10\,--\,17 & 0 \\
& S01-2 & 2011 Jan 14\,--\,24 & 3 \\
& S01-3 & 2011 Feb 16\,--\,28 & 9 \\
& S01 & outside campaign & 39 \\
& S02-1 & 2011 Dec 05\,--\,16 & 8 \\
& S02-2 & 2012 Jan 09\,--\,18 & 2 \\
& S02-3 & 2012 Jan 31\,--\,Feb 09 & 8 \\
& S02 & outside campaign & 24 \\
& S03-1 & 2012 Dec 04\,--\,14 & 1 \\
& S03-2 & 2013 Jan 08\,--\,18 & 1 \\
& S03-3 & 2013 Feb 10\,--\,17 & 3 \\\hline
& & total observations & 108 \\\hline
CIDA/Venezuela & S02-1 & 2011 Dec 05\,--\,16 & 5 \\
& S02-2 & 2012 Jan 09\,--\,18 & 6 \\
& S02-3 & 2012 Jan 31\,--\,Feb 09 & 8 \\\hline
& & total observations & 19 \\\hline
Rozhen/Bulgaria & S02-1 & 2011 Dec 05\,--\,16 & 5 \\
& S02-2 & 2012 Jan 09\,--\,18 & 3 \\
& S02-3 & 2012 Jan 31\,--\,Feb 09 & 0 \\\hline
& & total observations & 8 \\\hline
Xinglong/China & S03-1 & 2012 Dec 04\,--\,14 & 2 \\
& S03-2 & 2013 Jan 08\,--\,18 & 8 \\\hline
& & total observations & 10 \\\hline
Gunma/Japan & S00 & outside campaign & 4 \\\hline
\hline \hline
\end{tabular}
\\$^{\ast}$Name of the campaign runs e.g. S01-1: Season 1 (northern winter 2010/2011) Run 1
\end{table}

We first observed the 25\,Ori cluster from 2010 Jan to Apr using the 90\,cm Schmidt telescope of the University Observatory Jena. In this first phase, the Gunma Astronomical Observatory joined the photometric monitoring in 2010 Jan and Feb.\\ During the (northern) winter 2010/2011, 25\,Ori became a target of the YETI project where we arranged several international campaigns. The individual runs of these campaigns are typically 7 to 12 days long, and about three runs per year for three subsequent years. Since Orion is only observable in the winter half year from the northern hemisphere we divided the observations into seasons, where the start of the YETI monitoring (northern winter 2010/2011) corresponds to season 1 (S01). A summary of the participating observatories and their observations is given in Table~\ref{CCD_Kameras} and Table~\ref{Beobachtungen}.

\subsection{YETI monitoring}
\label{YETI_obs}

The complete YETI monitoring were done in $R$-band filter while the exposure times were chosen according to the individual telescope and instrumental setup. While most observations were obtained in the frame of the YETI campaigns, there are also some independent contributions, e.g. the  University Observatory Jena observed the cluster in every clear night in the four observing seasons. \\ At the University Observatory Jena we observed 25\,Ori in a total of 95 usable nights (108 nights in total see Table~\ref{Beobachtungen}), eight nights between 2010 Jan and Apr (S00), 43 nights between 2010 Oct and 2011 Apr (S01), 39 nights between 2011 Oct and 2012 Mar (S02), and five nights in 2012 Dec and 2013 Jan. The exposure time of the individual images was 50\,s. The University Observatory Jena thus accumulated 118.14\,h or 8506 individual exposures yielding photometric data reaching sufficient precision for the analysis of the fading events of CVSO\,30.\\ We observed 25\,Ori with the 1.5\,m reflector  of the Gunma Astronomical Observatory (GAO) using  the Gunma LOW-resolution Spectrograph \& imager (GLOWS) on 2010 Jan 29, 30, 31 and Feb 18; the exposure time was 30\,s during 2010 Jan and 60\,s on 2010 Feb 18. \\ In S02, the Centro de Investigaciones de Astronom\'{\i}a (CIDA) observed 25\,Ori with their 1\,m Coud\'{e} reflector which is equipped with an optical CCD camera. Because of the small FoV mosaicing was performed. The mean cadence of the 60\,s exposures was $\sim$8\,min which is sufficient to study variability on the timescale of $\sim$100\,min. In the campaign runs S02-1, S02-2, and S02-3, CIDA provided a total of 19 nights with usable data. \\ The Bulgarian National Astronomical Observatory (NAO), located in the Rhodopy Mountains at peak Rozhen, contributed five observing nights in S02-1 and three in S02-2. Their monitoring was carried out with their 50/70\,cm Schmidt and an exposure time of 60\,s. Usable data were obtained in seven out of eight nights. \\ The Xinglong station of the Beijing Astronomical Observatory (BAO) obtained 10 nights of data in two campaign runs in 2012 Dec (S03-01) and 2013 Jan (S03-02). With a 90\,cm Schmidt telescope \citep[60\,cm in Schmidt mode,][]{2007AJ....133.2061W} Xinglong collected 604 individual 60\,s exposures of CVSO\,30. \\ The photometric data were reduced following standard procedures including subtraction of bias (as overscan) and dark and dividing by a sky flat field. We calibrated the CCD images using the \begin{scriptsize}IRAF\end{scriptsize}\footnote{\begin{scriptsize}IRAF\end{scriptsize}
is distributed by the National Optical Astronomy Observatories, which are operated by the Association of Universities for Research in Astronomy, Inc., under cooperative agreement with the National Science Foundation.} routines textit{darkcombine}, \textit{flatcombine} and \textit{ccdproc}. In case of the  data of CIDA, Rozhen as well as Xinglong the basic data reduction was done by the observers.

\subsection{Follow-up photometry}
\label{OSN}

\begin{table}
\caption{Summary of the follow-up observations of CVSO\,30 done in the 2013/2014 and 2014/2015 seasons.}
\label{Follow-up}
\begin{tabular}{lcccc}
\hline \hline
Date & Observatory* & Filter & $N_{\mathrm{exp}}$ & $T_{\mathrm{exp}}$ (s)  \\\hline
2013 Nov. 01  & OSN-1.5m & $R$ & 106 & 60,120  \\
2013 Nov. 10  & OSN-1.5m &  $R$ & 98 & 120   \\
2013 Nov. 14  & OSN-0.9m &  $R$ & 118 & 120   \\
2013 Nov. 23  & OSN-0.9m &  $R$ & 120 & 120   \\
2013 Dec. 02  & OSN-1.5m &  $R$ & 182 & 60 \\
2013 Dec. 06  & OSN-1.5m &  $R$ & 220 & 60  \\
2013 Dec. 11  & OSN-1.5m &  $R$ & 157 & 60  \\
2013 Dec. 12  & Jena-0.6m &  $R$ & 41 & 180  \\\hline
2014 Nov. 27  & La Silla-2.2m &  $R$ & 60 & 10  \\
& &  $B$ & 60 & 30  \\
\hline \hline
\end{tabular}
\\
$^{\ast}$for a description see Table~\ref{CCD_Kameras}.
\end{table}

Following the end of international YETI campaign for 25\,Ori in 2013 Feb we obtained further photometric follow-up of CVSO\,30. To that end, we scheduled observations at the University Observatory Jena in the 2013/2014 season. Additionally, we were granted observation time at the 1.5\,m telescope of the Observatorio de Sierra Nevada in 2013 Oct-Dec, and at the 2.2-m MPG/ESO telescope at La Silla in 2014 Nov. The information about the follow-up observations are summarized in Table~\ref{Follow-up}. The basic data reduction was done as explained in \ref{YETI_obs}. \\ In 2013 we applied for observing time at the Observatorio de Sierra Nevada (OSN) operated by the Instituto de Astrof\'{i}sica de Andaluc\'{i}a, CSIC. We obtained a total of seven usable observations of the fading event. Five observations were obtained using the 1.5\,m reflector, additional two with the 90\,cm telescope. All observations were carried out in the $R$-band, and the exposure times varied between 60 to 120\,s; see Table~\ref{CCD_Kameras} for a summary of the equipment. \\ At the University Observatory Jena we observed CVSO\,30 in one more night in 2013 Dec. To achieve a better S/N for the $R$=15.2\,mag star the exposure time was set to 180\,s. The observations were done in the $R$ passband.\\ In the night 2014 November 26/27, we observed CVSO\,30 for $\sim$\,3\,hours with the WFI (Wide Field Imager) instrument mounted at the 2.2-m MPG/ESO telescope at La Silla. The observation covers a full fading event, including some pre- and post- brightness dip time. Individual exposures were obtained with two alternating filters: $BB\#B/123\_\mathrm{ESO}878$ and $BB\#R_{\mathrm{c}}/162\_\mathrm{ESO}844$ (hereafter $B$ and $R$ filters) so that we essentially obtained simultaneous two-band photometry. The integration time was 30\,s for the $B$- and 10\,s for the $R$ filter. To minimize overheads, the filters were only changed after two subsequent exposures. The WFI detector consists of eight CCD chips. The source was located on chip no.~51\footnote{see http://www.eso.org/sci/facilities/lasilla/instruments/wfi/overview.html for details} for all but the first two exposures, which we therefore, excluded from the following analysis.

\section{Photometry}
\label{photometry}

While the basic data reduction was either done by the individual observers or in Jena, the photometry was carried out uniformly for all observations. Magnitudes were derived by performing aperture photometry with the dedicated \begin{scriptsize}IRAF\end{scriptsize} user script \textit{chphot} which is based on the standard \begin{scriptsize}IRAF\end{scriptsize} routine \textit{phot}. Our script allows us to obtain simultaneous photometry of all field stars in an image. For this purpose a list of the pixel coordinates of all detectable stars was created using \begin{scriptsize}SOURCE EXTRACTOR\end{scriptsize} \citep[\begin{scriptsize}SEXTRACTOR\end{scriptsize};][]{1996A&AS..117..393B}. Our final target list is based on the maximum FoV of all telescopes participating in any YETI campaign (i.e., 2.7$^{\circ}$\,x\,2.7$^{\circ}$) and contains a total of 30894 stars.\\ The positions of the stars on the individual images were determined using either \begin{scriptsize}ECLIPSE\end{scriptsize} Jitter \citep{1997Msngr..87...19D} or \begin{scriptsize}SEXTRACTOR\end{scriptsize}. The result was compared with the reference list of stars as done by \citet{2014AN....335..345E}. While \begin{scriptsize}ECLIPSE\end{scriptsize} Jitter is faster, it is only applicable to image time series with small pixel shifts. \begin{scriptsize}SEXTRACTOR\end{scriptsize} can also handle large pixel shifts, which are present, e.g., in the case of mosaicing.\\ To transfer the reference list of stars from one telescope to the other we used either simple coordinate transformation based on pixel scale and orientation of the detector (only possible for no or little field distortions i.e. small FoVs) or we did an object detection on a astrometric calibrated image and matched the right ascension and declination with \begin{scriptsize}TOPCAT\end{scriptsize} \citep{2005ASPC..347...29T} to the initial target list. In the latter case the images were calibrated using \textit{astrometry.net} \citep{2010AJ....139.1782L}.\\ Differential photometry was performed with the \begin{scriptsize}PHOTOMETRY\end{scriptsize} program developed by \citet{2005AN....326..134B}. By taking a weighted average of all available field stars, the program creates an optimised artificial comparison star. The individual weights are deduced from the standard deviation of the processed time series for each star. Faint and/or variable stars with a high standard deviation enter the artificial comparison star with a low weight while bright, constant stars contribute with higher weights. With this method, LCs for all field stars can be obtained by comparison with the artificial standard star.\\ To find the optimal radius for the extraction aperture, ten different aperture radii were tested; the annulus for sky subtraction remained fixed. For each aperture we determined the standard deviation of the LCs for a sample of constant stars which were selected as the ones with the highest weights in the artificial comparison star. The  radius that yields the smallest sum of standard deviations was finally chosen.\\ The photometry was obtained for individual nights except for the observations with CIDA for which the number of images per night remains small due to the mosaicing. To combine data from different nights we applied the procedure described in \citet{2014AN....335..345E} which is based on the night-to-night difference in the differential magnitudes of constant standard stars.\\ To account for systematic effects that are present in the LCs of all stars, two more steps were carried out. First, the LC of a bright, constant star was used to identify outliers attributable, e.g., to dome vignetting, changing weather conditions or jumps due to large pixel shifts. In particular, we applied sigma-clipping to the LC of the chosen reference star and removed the identified outliers from all LCs. Second, we calculated the average photometric error of CVSO\,30 and removed every data point whose uncertainty exceeds twice the mean value. A factor of two was found appropriate to keep LCs showing reasonable behaviour, but eliminate obviously inappropriate sections of the LC caused, e.g., by non-optimal weather conditions.

\section{Light curve analysis}

The original LCs for all usable nights for each telescope are shown in the appendix Figs.~\ref{LC_S00}, \ref{LC_S01_50s}, \ref{LC_S02_GSH}, \ref{LC_S03_GSH}, \ref{LC_S00_Gunma}, \ref{LC_S02_CIDA}, \ref{LC_S02_Rozhen}, and \ref{LC_S03_Xinglong}. The expected time windows of the fading events are highlighted as grey shaded areas. The mid-times were calculated using the updated ephemeris (see section \ref{Transit_timing}),and the duration was fixed to the value of \citet{2012ApJ...755...42V}.\\ The LCs of the Gunma Astronomical Observatory in S00 provide partial coverage of a single fading event. While the associated LC does show an increase in brightness at the expected time of egress, no pre-event reference level is available. Therefore we cannot exclude that the change in brightness is due to stellar variability.\\ During the first season of the international campaign (S01), the observations at the University Observatory Jena covered two clearly detected fading events (JD 2455534 and 2455601). Partial coverage of another two fading events (JD 2455615 and 2455619) shows a clear egress, however, the ingress remained unobserved. In a few more cases a detection of the fading event is evident but remains insignificant due to larger measurement uncertainties or data gaps (JD 2455533, 2455584, 2455614, 2455618). Interestingly, no brightness dip is visible during the fully covered window at JD 2455627. In addition to the stellar variability probably attributable to rotation, a few likely flares at JD 2455478, 2455614 (during the fading event), 2455628 and probably at 2455649 (also during the fading event) can be identified.\\ In S02 altogether 27 individual fading events were at least partially covered by our observations at Jena, CIDA and Rozhen. Unfortunately, observational overlap is only available for the night at JD 2455941 during which the Jena and Rozhen observatories simultaneously provide post-event coverage. While no fading event was observed simultaneously, the observed stellar variability is consistent in time and amplitude for both observatories so that we can exclude systematic effects between the telescopes.\\ In the last observing season, S03, we collected two complete and two partial fading events. The depth of the fading event at JD 2456305 seems to be smaller than other fading events observed in previous seasons.\\ While the fading events show a varying depth, several events seem to have asymmetric profiles (e.g. JD 2455890 or 2455944). In general, it is evident, even though not significant, that the shape changes between different fading events. Interestingly, there are as few as eight days between a clear detection and a non-detection (e.g. JD 2455619 and 2455627). \\ The original LCs of our follow-up observations are shown in the appendix Fig.~\ref{LC_CVSO30_Follow_up}. For the OSN data we detected a clear signal of the fading event in all seven observations. The analysis of the data revealed that the beginning of the fading event was missed by applying the ephemeris given by \citet{2012ApJ...755...42V}. Therefore the ephemeris were refined. The LC from JD 2456633 shows a feature that could be attributable to a flare during the egress at the end of the fading event. Besides these occasional events during the fading event, the shape seems to change. While JD 2456598, 2456607 and 2456629 look ``u''-shaped, the remaining fading events are more similar to a ``v''. Also the follow-up LC of the University Observatory Jena shows a clear detection of the brightness dip. Interestingly, all events observed at the OSN and the University Observatory Jena in the 2013/2014 season seem to be shallower compared to the fading events from S02. \\ The LCs obtained at the 2.2-m MPG/ESO telescope at La Silla in $B$- and $R$-band are shown in Fig.~\ref{2456988}. According to our modelling (see following section), the LCs show a fading event with the same depth ($\sim$\,3 per cent) and shape in both bands. The S/N (depth of the fading event divided by the standard deviation of the out-of-event data points) is 2.7 and 2.4 for the $B$- and $R$-band LC, respectively. Although the detection of the same profile of the fading event in two photometric bands lends some support to the planetary transit hypothesis because a starspot is expected to produce a colour-dependent signal, the simultaneous multiband observations by \citet{2015arXiv150902176Y} show a deeper brightness dip in the bluer band in four out of five cases. Both these observations and the fast evolution observed by us are hard to reconcile with the planetary hypothesis.\\ Stellar activity is a problem for the detection of the transit events and the derivation of the planetary properties. As expected for a T-Tauri star the LCs of the candidate are dominated by stellar variability. In order to model the fading event for a derivation of the planetary candidate properties, the effects of the stellar variability in the LCs need to be minimized. As shown by \citet{2015MNRAS.450.3991K} CVSO\,30 exhibits a complicated non-sinusoidal quasi-periodicity with several frequencies which are, so far, not well understood. Therefore it is not possible to create a full parametric model for the stellar variability. Thus, we decided to treat every LC individually with a model that best fits the data. The LC were detrended by fitting either a polynomial up to the third order or a spline to the out-of-event points. The \begin{scriptsize}IDL\end{scriptsize} routines \textit{splinesm} and \textit{POLY\_FIT} were used for this purpose. This method generated decent detrended LCs for complete fading events with data point of both sides of the event. Since it is not possible to extrapolate the behaviour of the variability to the other side in case of partial fading events, the detrended LCs are not reliable. For example, a change of the polynomial order could result in a different  depth. Therefore, the results obtained with the partial LCs should be taken with caution.

\subsection{Transit Fitting}
\label{Transit_Fitting}

\begin{table}
\centering
\caption{Results of the LC modelling and derived physical properties of the CVSO\,30 system. Values given in \citet{2012ApJ...755...42V} are shown for comparison.}
\label{phys_prop_CVSO30}
\renewcommand{\arraystretch}{1.1}
\begin{tabular}{lr@{\,$\pm$\,}lr@{\,$\pm$\,}l}
\hline \hline
 Parameter & \multicolumn{2}{c}{This work} & \multicolumn{2}{c}{\citet{2012ApJ...755...42V}} \\ \hline \hline
& \multicolumn{4}{c}{Measured} \\ \hline 
$\frac{a}{R_{\ast}}$ & 1.805 &   $\mathbf{^{0.004}_{0.008}}$ & 1.685 & 0.064 \\
$k=\frac{R_{\mathrm{Pl}}}{R_{\ast}}$ & 0.1916 & $\mathbf{^{0.0025}_{0.0040}}$ & 0.1838 & 0.0097 \\
$u$ (\textit{R} band) & 0.51 & $\mathbf{^{0.05}_{0.06}}$ & \multicolumn{2}{c}{} \\
$v$ (\textit{R} band) & 0.47 & $\mathbf{^{0.06}_{0.06}}$ & \multicolumn{2}{c}{} \\
$u$ (\textit{B} band) & 0.40 & $\mathbf{^{0.30}_{0.24}}$ & \multicolumn{2}{c}{} \\
$v$ (\textit{B} band) & 0.20 & $\mathbf{^{0.31}_{0.31}}$ & \multicolumn{2}{c}{} \\ \hline
& \multicolumn{4}{c}{Derived} \\ \hline 
$a$  [au] & 0.00840 & 0.00036$^{a}$ & 0.00838 & 0.00072 \\
$R_{\mathrm{Pl}}$  [R$_{\mathrm{Jup}}$] & 1.87 & $\mathbf{^{0.08}_{0.09}}$ & 1.91 & 0.21  \\
$\rho_{\mathrm{Pl}}$  [$\mathrm{\rho}_{\mathrm{Jup}}$] & 0.46 & $\mathbf{^{0.07}_{0.07}}$ &  \multicolumn{2}{c}{}   \\
$T_{\mathrm{eq}}$ [K] & \multicolumn{2}{c}{1826$^{b}$} & \multicolumn{2}{c}{}   \\ 
$\Theta$ & 0.074 & $\mathbf{^{0.010}_{0.010}}$ & \multicolumn{2}{c}{}  \\ 
$R_{\mathrm{A}}$  [R$_{\mathrm{\odot}}$] & 1.00  & $\mathbf{^{0.04}_{0.04}}$ & 1.07 & 0.10  \\
$\rho_{\mathrm{A}}$  [$\mathrm{\rho}_{\mathrm{\odot}}$] & 0.39 & $\mathbf{^{0.002}_{0.005}}$ & \multicolumn{2}{c}{}  \\ \hline
\end{tabular}
\\
$^{a}$ calculated with the updated period (see section \ref{Transit_timing})\\
$^{b}$ No uncertainty given for the stellar $T_{\mathrm{eff}}$ in \citet{2005AJ....129..907B}
\end{table}

The detrended LCs were modelled with the Transit Analysis Package\footnote{http://ifa.hawaii.edu/users/zgazak/IfA/\begin{scriptsize}TAP\end{scriptsize}.html} \citep[\begin{scriptsize}TAP\end{scriptsize} v2.1,][]{2012AdAst2012E..30G}. Using Bayesian Markov Chain Monte Carlo (MCMC) techniques \begin{scriptsize}TAP\end{scriptsize} fits the analytic transit model of \citet{2002ApJ...580L.171M} (with quadratic limb darkening) to the LCs with the fast exoplanetary fitting code \citep[\begin{scriptsize}EXOFAST\end{scriptsize},][]{2013PASP..125...83E}. To determine parameter uncertainties \begin{scriptsize}TAP\end{scriptsize} incorporates a wavelet based likelihood function developed by \citet{2009ApJ...704...51C}. Since this technique parametrizes uncorrelated as well as time-correlated noise it allows one to estimate robust parameter uncertainties.\\ During our observations we detected 38 fading events with highly variable quality that varies in shape between different observations. The shape of a transit is described by the depth, the total duration and the duration of ingress and egress. Both, the total duration and the depth of a transit, depend on the orbital inclination $i$.  \\ Using the assumption of a precessing planetary orbit \citep{2013ApJ...774...53B} we could facilitate the modelling. We account for the changing shape by fitting an individual inclination for each fading event, while the semimajor-axis scaled by stellar radius $\frac{a}{R_{\ast}}$, the planetary to stellar radii ratio $k$, and the LD coefficients (only $R$-band) were linked together for all LCs. The orbital period $P$, the eccentricity $e$ and the argument of periastron $\omega$ were kept fixed while the mid-times of the event $T_{\mathrm{c}}$, and the inclination were allowed to vary separately. To avoid nonphysical results that disagree with the RVs measured by \citet{2012ApJ...755...42V} we constrain the fitting parameters. The inclination was chosen to vary between 56$^{\circ}$ \citep[minimum inclination for a transit to occur for an orbital period of $\sim$0.44\,d and a stellar radius of $\sim$1\,R$_{\odot}$,][]{2012ApJ...755...42V} and 90$^{\circ}$. Using the mass-radius relation of young stars by \citet{2015arXiv150502446K} and the results of \citet{2005AJ....129..907B} and \citet{2012ApJ...755...42V}, the range of $\frac{a}{R_{\ast}}$ was set to 1.29 to 1.81. While evolutionary models for irradiated planets by \citet{2003A&A...402..701B} predicts that CVSO\,30\,b cannot be smaller than 1.6\,R$_{\mathrm{Jup}}$, the size of the Roche lobe (calculated from the mass ratio and the semimajor axis) sets an upper limit of 1.9\,R$_{\mathrm{Jup}}$. Therefore $k$ was allowed to vary between 0.117 and 0.195. The global results of the LC modelling are summarized in Table~\ref{phys_prop_CVSO30}, while the individual values for the mid-time and the inclination are given in the appendix (Table~\ref{CVSO30_results}).

\begin{figure*}
 \centering
  \includegraphics[width=0.95\textheight, angle=270]{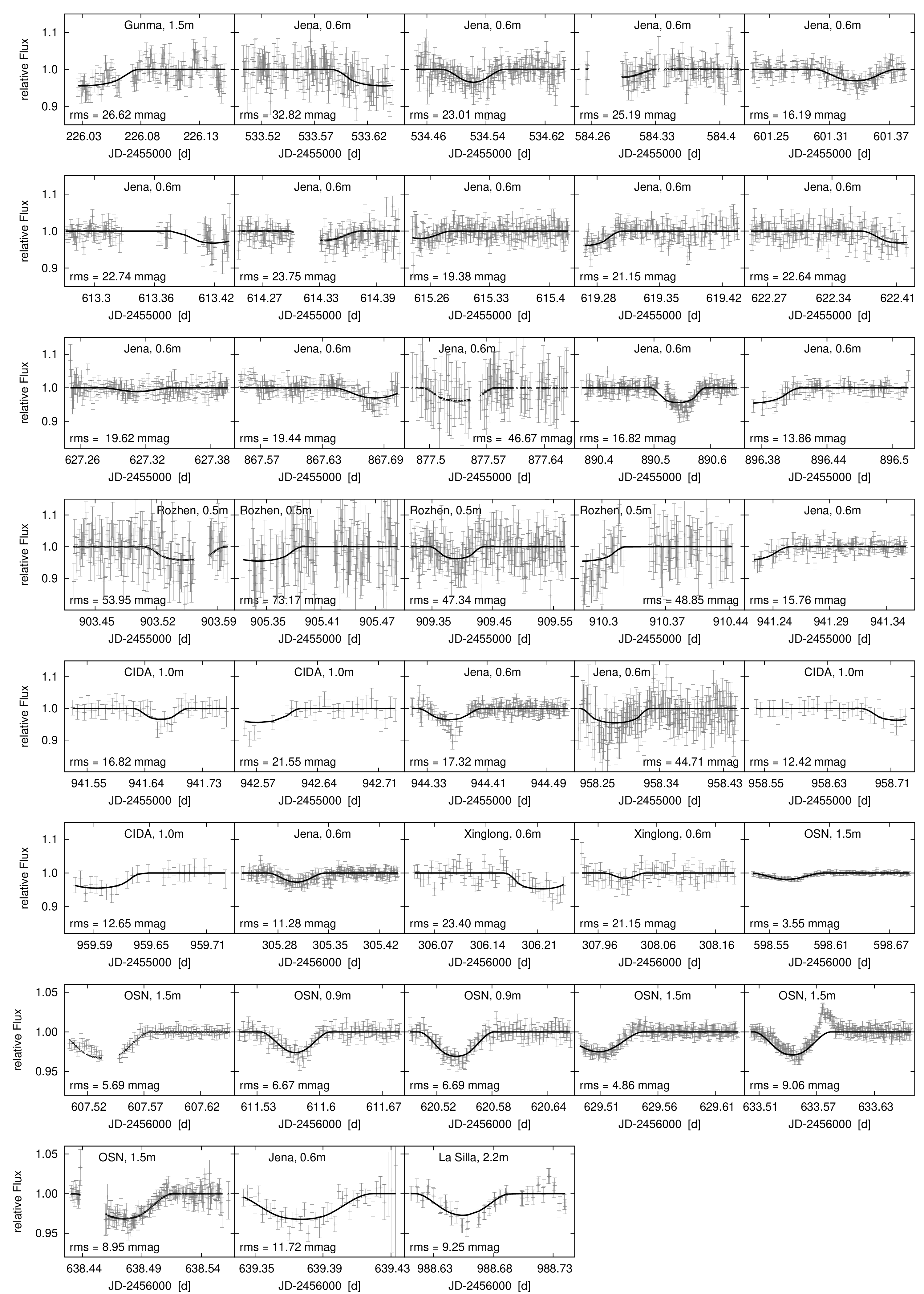}
  \caption{All observed $R$-band fading events of CVSO\,30 with the best-fitting model shown as black solid line (or as individual points for LCs with large gaps). The observatory, telescope size, and the $rms$ of the fit are indicated in each individual panel.}
\label{Transits_all}
\end{figure*}

\begin{figure}
  \centering
  \includegraphics[width=0.32\textwidth,angle=270]{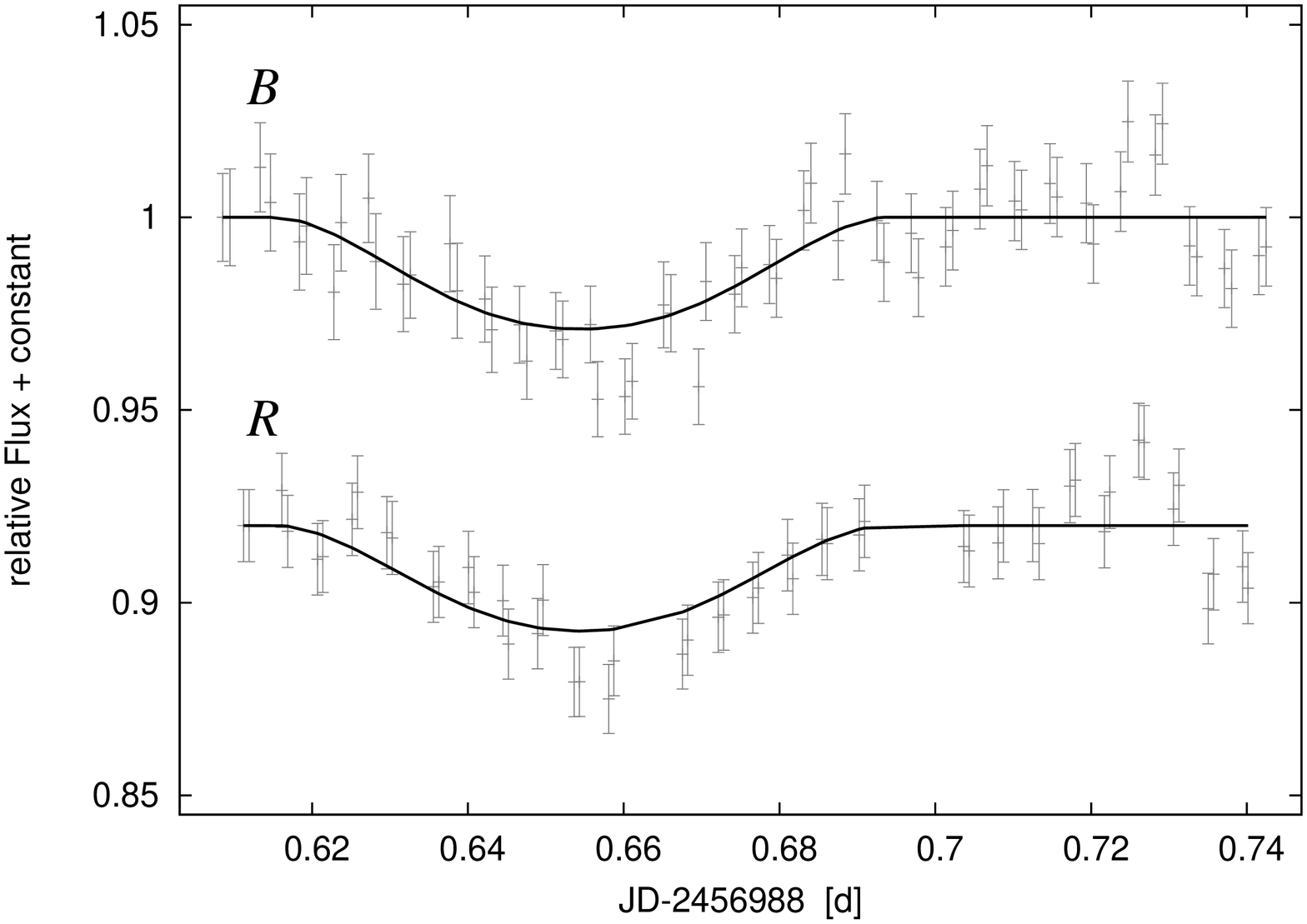}
  \caption{$B$- and $R$-band LCs of JD 2456988 observed at the MPG/ESO 2.2\,m Telescope at La Silla. The fading event is clearly detected and shows the same depth ($\sim$\,3 per cent) in both wavelength bands.}
  \label{2456988}
\end{figure}

\subsection{Physical properties}

The results of the LC modelling allow us to calculate stellar, planetary, and geometrical parameters. \\ The mean stellar density $\rho_{\ast}$ can be derived directly from the parameters obtained from the LC modelling using
\begin{equation}
\label{density}
\rho_{\ast}=\frac{3\mathrm{\pi}}{GP^{2}}\left( \frac{a}{R_{\ast}}\right)^{3}
\end{equation}
\citep{2010exop.book...55W}, where $G$ is the gravitational constant. To calculate the semimajor axis $a$, we inserted the improved period (see Section \ref{Transit_timing}) into Kepler's third law. By resolving $a$/$R_{\ast}$ and $R_{\mathrm{Pl}}$/$R_{\ast}$ using the already determined value for $a$ we deduce values for $R_{\ast}$ and $R_{\mathrm{Pl}}$. The radius of the planet candidate as well as its mass were used to calculate the density $\rho_{\mathrm{Pl}}$. The impact parameter $b$ and the depth were calculated for each individual fading event.  \\ In addition, we calculated the equilibrium temperature of the planet candidate $T_{\mathrm{eq}}$ \citep[assuming a Bond albedo\,=\,0 and only little energy redistribution across the surface of the planet candidate;][]{2007ApJ...671..861H} and the Safronov number $\Theta$ \citep{1972epcf.book.....S}, the square of the ratio of escape velocity of the planet candidate $v_{\mathrm{esc}}$ and orbital velocity $v_{\mathrm{orb}}$. The results of the calculations for the global parameters are summarized in Table~\ref{phys_prop_CVSO30}, while the individual impact parameters and depths are given in Table~\ref{CVSO30_results} in the appendix. In general, our derived system parameters are slightly smaller but consistent with \citet{2012ApJ...755...42V} within the error bars.

\subsection{Timing analysis}
\label{Transit_timing}

\begin{figure}
  \centering
  \includegraphics[width=0.32\textwidth,angle=270]{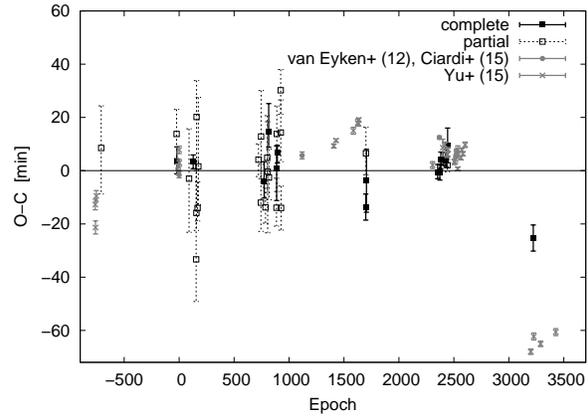}
  \caption{The O--C-diagram of CVSO\,30\,b. The black filled and open (with dashed error bars) symbols denote the complete and the partial fading events, respectively. The data points from \citet{2012ApJ...755...42V}, \citet{2015arXiv150608719C} and \citet{2015arXiv150902176Y} are shown in grey. The solid line represents the updated ephemeris given in equation (\ref{Elemente_CVSO30}).}
  \label{B_R}
\end{figure}

The mid-times of the fading event that were obtained by modelling (see section \ref{Transit_Fitting}) are given in simple Julian date (JD). They have been converted into the barycentric Julian Date based on the barycentric dynamic time (BJD$_{\mathrm{TDB}}$) using the online converter\footnote{http://astroutils.astronomy.ohio-state.edu/time/utc2bjd.html} by \citet{2010PASP..122..935E}. While we excluded four LCs with no clear detection of the brightness dip from the timing analysis, we included the the original published mid-time at epoch zero computed from many individual fading events \citep{2012ApJ...755...42V}. In total we used 39 mid-times to compute updated ephemeris using an error weighted linear fit. The result is given in equation (\ref{Elemente_CVSO30}), where $E$ denotes the epoch: 
\begin{equation}
\label{Elemente_CVSO30}
\begin{array}{r@{.}lcr@{.}l}
T_{\mathrm{c[BJD_{TDB}]}}(E)=(2455543 & 9420 & + & E\cdot 0 & 4483973)\,\mathrm{d} \\
\pm0 & 0007 &  & \pm0 & 0000004
\end{array}
\end{equation}
The period of the fading event $P$ is 1.36\,s shorter and 100 times more precise than the one given in \citet{2012ApJ...755...42V}. \\ Our updated ephemeris were used to calculate the ``observed minus calculated'' (O--C) residuals. The obtained O--C values are listed in Table~\ref{CVSO30_results} in the appendix. The resulting O--C diagram is shown in Fig.~\ref{B_R}. We also included the published mid-times of \citet{2015arXiv150608719C} and \citet{2015arXiv150902176Y}. Significant deviations, up to 5.3$\sigma$ from the ``zero''-line, can be seen in the O--C diagram. The reduced $\chi^{2}$ for the error weighted linear fit is 2.3 and is a result of the large scatter ($\sim$20\,min only for the sample of completely covered events) of the O--C values within each season.\\ \citet{2015arXiv150902176Y} claimed from their timing analysis that the fading events are not strictly periodic and reported on a steady decrease in the period. While our mid-times are in very good agreement with \citet{2015arXiv150902176Y} in the 2010/2011 and 2013/2014 seasons they do not coincide in the  2012/2013 and 2014/2015. Our data may suggest that the mid-times deviate from the strictly periodic case but we cannot confirm the fast orbital decay estimated by \citet{2015arXiv150902176Y}. Moreover, \citet[][conference poster]{2015ESS.....312004Y} reported one more event observed around epoch $\sim$4000 which do not support the fast period change. \\ A generalised Lomb--Scargle periodogram \citep[\begin{scriptsize}GLS\end{scriptsize};][]{2009A&A...496..577Z} of our O--C values of the complete covered fading events was computed to search for a periodicity. The highest peak $P_{\mathrm{TTV}}$\,=\,187.5\,$\pm$\,0.9\,epochs, amplitude 9.8\,$\pm$\,1.5\,min) in the periodogram corresponds to a False-Alarm probability (FAP) of 54.3 per cent. Hence, our data do not show any significant periodicity in the mid-times of the fading events. Non-periodic timing variations may still be possible. But with the varying quality of our data set systematic errors in the mid-times that result from the detrending of the LC cannot be excluded.  
 
\subsection{Orbital precession}

\begin{figure}
  \centering
  \includegraphics[width=0.32\textwidth,angle=270]{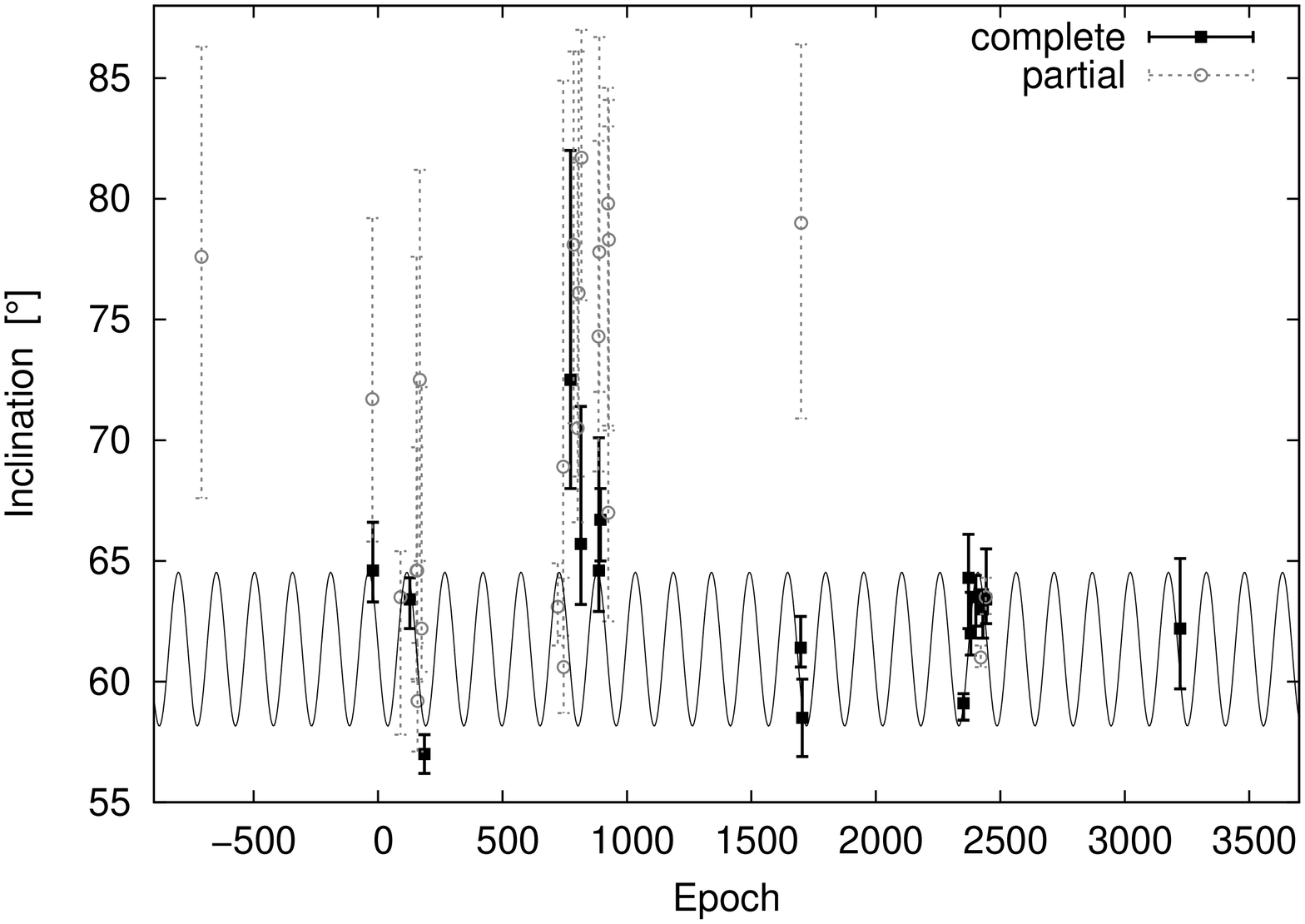}
  \caption{Inclination vs. epoch for CVSO\,30\,b. The black filled and grey open (with dashed error bars) symbols denote the complete and the partial fading events, respectively. The sinusoidal curves corresponds to the period found in the Lomb--Scargle periodogram (see Fig.~\ref{GLS_Incl}) }
  \label{Epoch_Incl}

  \centering
  \includegraphics[width=0.32\textwidth,angle=270]{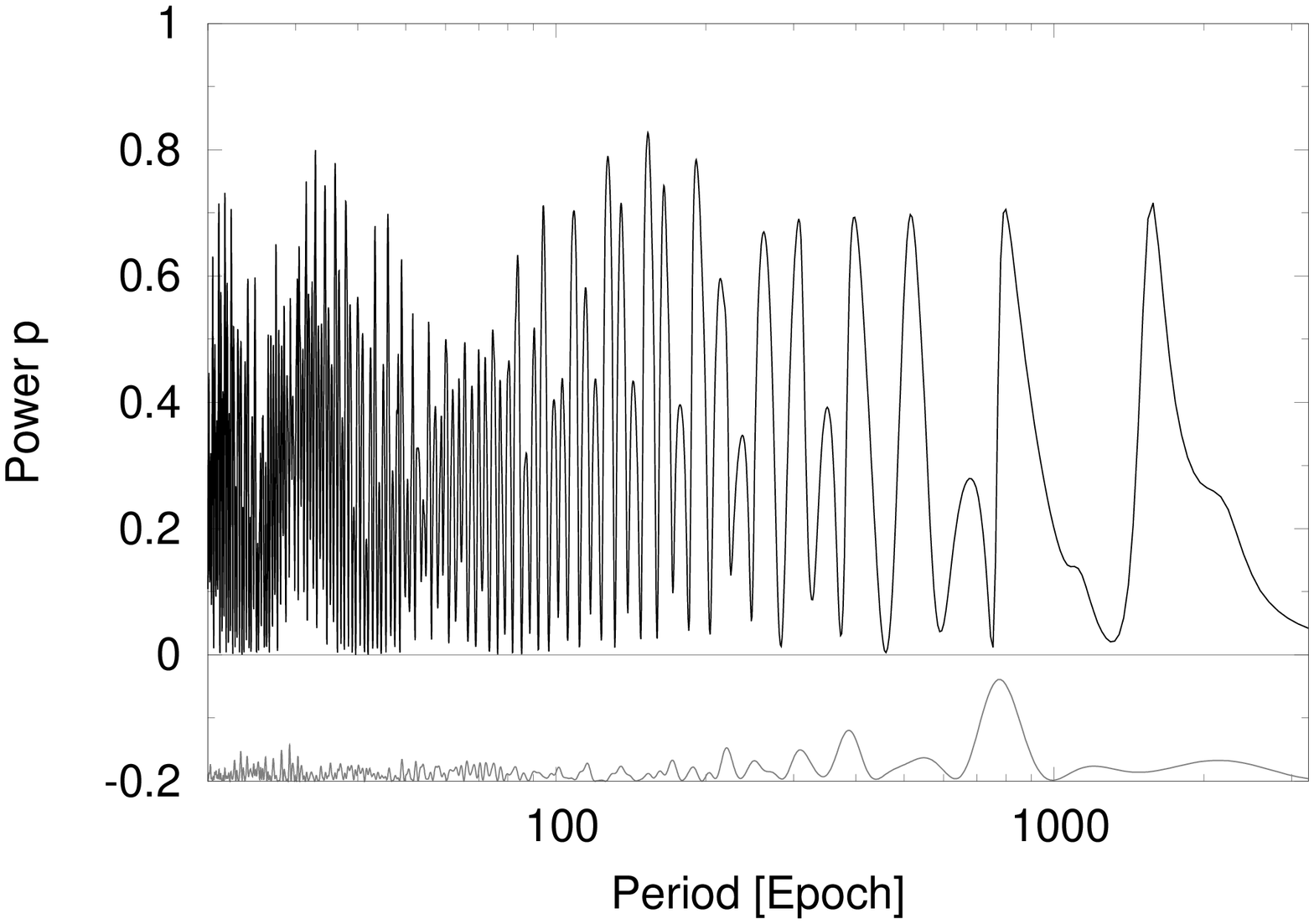}
  \caption{Generalised Lomb--Scargle periodogram (top panel) and window function (bottom panel) for the Inclination of CVSO\,30\,b. The highest peak corresponds to $P_{\mathrm{Incl}}$\,=\,152.97\,$\pm$\,0.55\,epochs ($\sim$\,68.5\,d)}
  \label{GLS_Incl}
\end{figure}

As explained in section \ref{Transit_Fitting} we accounted for the changing depth and duration by fitting an individual inclination for each fading event.  Any orbital precession should be seen in the change of the inclination. Therefore we plotted the inclination over the epoch and carried out a period search using \begin{scriptsize}GLS\end{scriptsize}. The periodogram for the 16 complete fading events with a period range of 20\,d (Nyquist frequency) to 3243\,d (longest baseline) is shown in Fig.~\ref{GLS_Incl}. Although \begin{scriptsize}GLS\end{scriptsize} outputs a FAP for the highest peak at $P_{\mathrm{Incl}}$\,=\,152.97\,$\pm$\,0.55\,epochs ($\sim$\,68.5\,d) of 0.2 per cent, we will not claim a significant detection considering the quality and cadence of the data. Our best fitting  period is smaller than the previously published precession periods that were derived from numerical models. However, taking into account that between a clear detection (JD 2455619) and a non-detection (JD 2455627) there are as few as 8\,d a lower precession period seems to be plausible.

\section{Stellar Rotation}
\label{stellar_rotation}

\begin{figure}
  \centering
  \includegraphics[width=0.32\textwidth,angle=270]{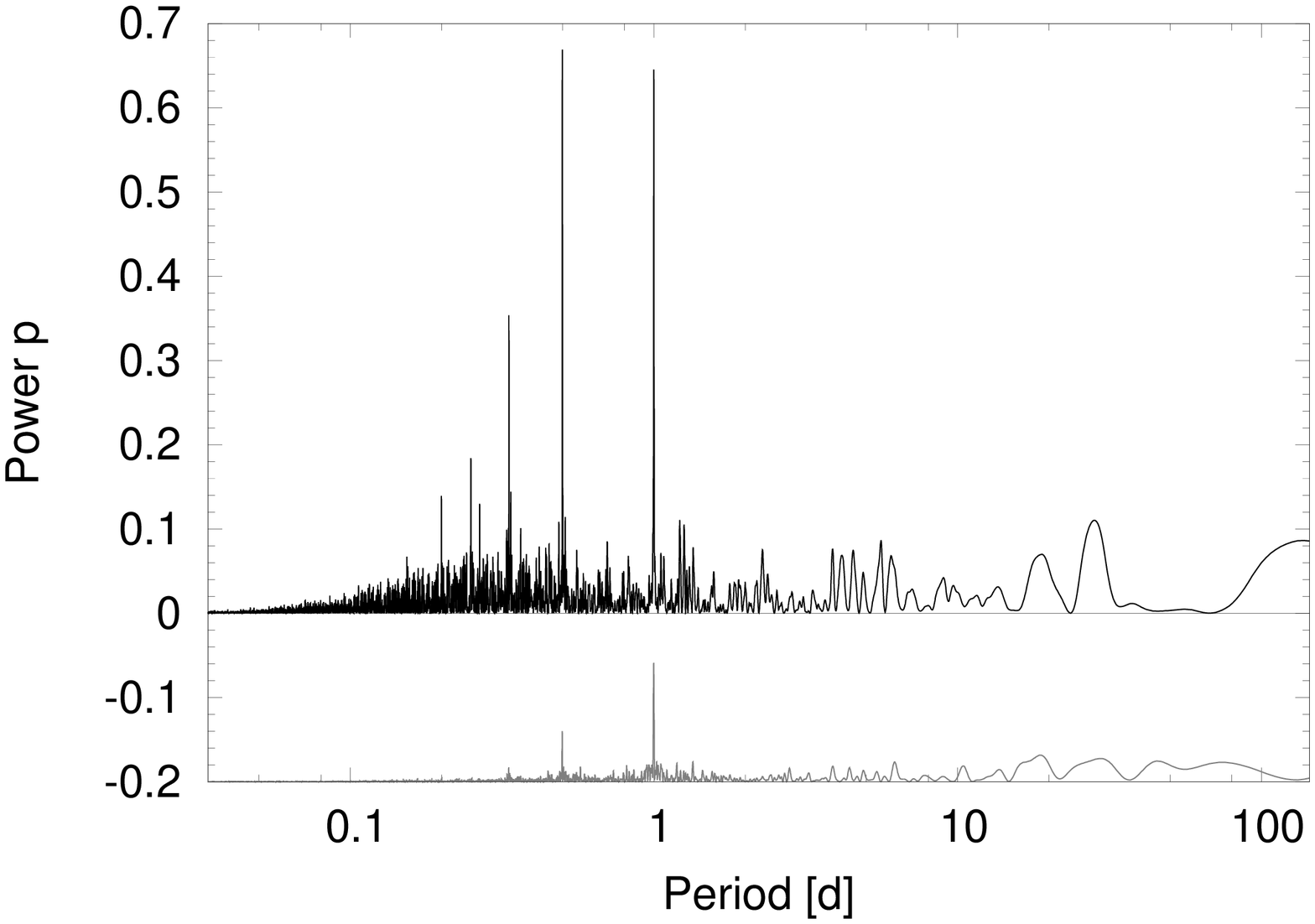}
  \caption{Generalised Lomb--Scargle periodogram (top panel) and window function (bottom panel) for the S02 data of the University Observatory Jena, CIDA and Rozhen (data inside the fading events were removed). The highest peak with a period of $P_{\mathrm{rot}}$\,=\,0.49927\,$\pm$\,0.00001\,d. No peak is seen at the previously reported rotation period $P_{\mathrm{rot}}$\,=\,0.4484\,d (corresponding power: p\,=\,0.0165).}
  \label{GLS_rotation}

  \centering
  \includegraphics[width=0.32\textwidth,angle=270]{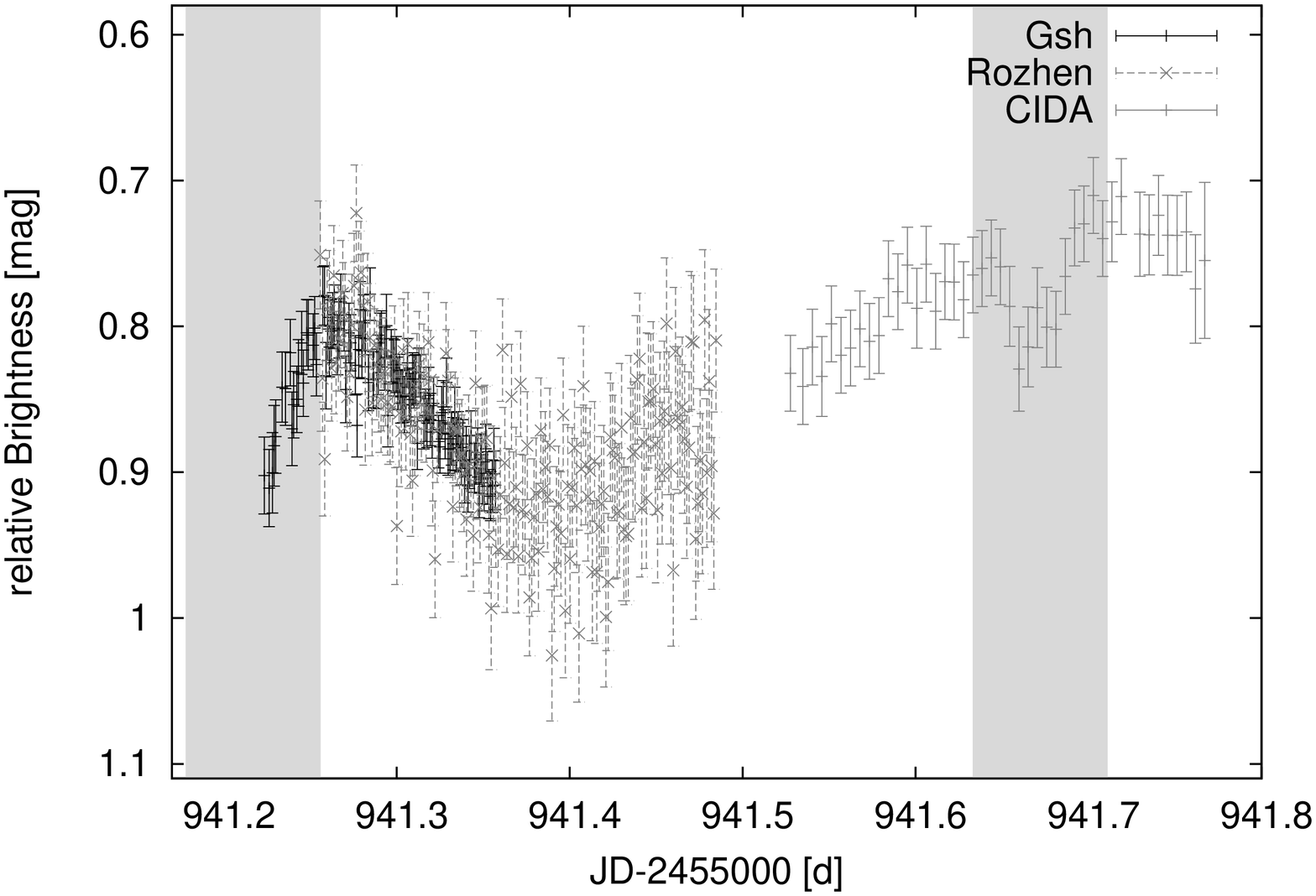}
  \caption{Our best covered night JD 2455941 in S02 which was used for the determination of the stellar rotation period. The observations of CIDA/Venezuela are complementary to the other two observatories and help to cover a complete phase.}
  \label{2455941_all}
\end{figure}

Understanding the stellar variability is of critical importance for the investigation of the system properties. Due to stellar rotation and spots on the surface, CVSO\,30 shows a quasi-periodic variation. Additionally it also shows irregular variability, i.e. in the form of occasional flares. \citet{2012ApJ...755...42V} investigated their two years' data and found a strong peak at $\sim$0.448\,d which matches the orbital period of the planet candidate. They concluded that the star is co-rotating or in near co-rotation with the planetary orbit. They also mentioned another peak at $\sim$0.998\,d but claimed that this is probably a result of the observing cadence. However, since the planetary orbit is mis-aligned to the stellar rotation axis by $\sim70^{\circ}$, \citet{2013ApJ...774...53B} as well as \citet{2015arXiv150604829K} stated that synchronous stellar rotation is almost impossible to achieve via tidal torques. \\ \citet{2015MNRAS.450.3991K} reanalysed the data set of \citet{2012ApJ...755...42V} along with their own six nights of observation in order to investigate the stellar rotation. He found two fundamental periods, 0.33\,d (or it's alias of 0.50\,d) and 0.448\,d with amplitudes varying from 25-43\,mmag. \\ With our 4 years baseline of 25\,Ori monitoring we carried out a period analysis. Since the observing gap between two consecutive seasons is quite long and the amplitude of variation changes over the years we analysed the data on a seasonal basis. With \begin{scriptsize}GLS\end{scriptsize} we computed a Lomb--Scargle periodogram for every season and different combinations of telescopes individually. The data inside the time windows for the fading event were removed before the analysis. A visual inspection of the LCs in consecutive nights revealed the same behaviour, hence the variation period should be $\sim$1\,d or a fraction or multiple of that. This is also confirmed by the better quality LCs of \citet{2015arXiv150902176Y}. From the change of the position of the brightness dip relative to the maxima or minima of the overall light variation on consecutive days the rotation period has to be longer than the period of the fading events \citep[][e.g. their LCs from 2010 Dec 9 and 10]{2015arXiv150902176Y}. Fig.~\ref{GLS_rotation} shows a typical Lomb--Scargle periodogram  for S02. Table~\ref{rotation_period} summarizes the strongest peak for all obtained Lomb--Scargle periodogram. We also included a single computation for our best covered night (JD 2455941, see Fig.~\ref{2455941_all}) and our best covered week (JD 2455958--2455967). In all cases we found that the rotation period of the star is $\sim$0.5\,d which is slightly larger than the orbital period of the planet candidate. This agrees with the results of \citet{2015ApJ...807...78F} who found that active host stars with big close-in companions tend to have rotational periods larger than the orbital periods of their companions. Our rotation period is confirmed by \citet{2016AAS...22714510T} who used several methods to determine the most probable period for 1974 confirmed T Tauri members of various sub-regions of the Orion OB1 association using a much larger data set including the YETI data. It is also consistent with the main frequency reported by \citet{2015MNRAS.450.3991K} ($f_{1}\approx3$ or its alias $2d^{-1}$). We can not find the previously reported rotation period $\sim$0.448\,d. As expected, the amplitude of the variability changes in the different seasons between $\sim$30 and $\sim$100\,mmag. Interestingly, the obtained periods do not agree within the error bars. Even if using the data from the same season but with different telescopes we found a deviation in period and amplitude. This could be a bias of the rotation period itself. Since the star rotates with $\sim$0.5\,d we always see the same phase with only one telescope. Only after adding data of another telescope at a very different longitude can we cover a full phase. The data from S00 of Jena/Germany and Gunma/Japan are, for example, complementary with no overlap in phase. Therefore the rotation period of CVSO\,30 is an excellent example for the importance of global telescope networks such as YETI.

\begin{table}
\centering
\caption{Summary of the strongest peak in the Lomb--Scargle periodogram that was obtained by period analysis for every season and different combinations of telescopes.}
\label{rotation_period}
\begin{tabular}{lccc}
\hline \hline
Season & Observatory* & Period & Amplitude \\ 
& & (d) & (mmag) \\\hline \hline
S00 & Jena, Gunma & 0.49932 $\pm$ 0.00007 & 52.7 $\pm$ 1.8 \\
S01 & Jena & 0.49936 $\pm$ 0.00001 & 41.0 $\pm$ 0.8 \\
S02 & Jena & 0.49945 $\pm$ 0.00001 & 55.3 $\pm$ 0.5 \\
\multirow{2}{*}{S02} & Jena, CIDA, &  \multirow{2}{*}{0.49927 $\pm$ 0.00001} & \multirow{2}{*}{78.6 $\pm$ 0.6}  \\
 & Rozhen &  &   \\
S03 & Jena & 0.49899 $\pm$ 0.00005 & 38.0 $\pm$ 1.0 \\ 
S03 & Xinglong & 0.49928 $\pm$ 0.00022 & 28.9 $\pm$ 1.8 \\
\multirow{2}{*}{S03} & Jena, & \multirow{2}{*}{0.49896 $\pm$ 0.00005} & \multirow{2}{*}{37.3 $\pm$ 1.0} \\
 & Xinglong &  &   \\\hline
\multirow{2}{*}{2455941} & Jena, CIDA, & \multirow{2}{*}{0.53834 $\pm$  0.00652} & \multirow{2}{*}{99.3 $\pm$  2.1} \\
 & Rozhen &   &  \\
2455958- & \multirow{2}{*}{Jena, CIDA} & \multirow{2}{*}{0.49745 $\pm$  0.00036} & \multirow{2}{*}{72.7 $\pm$  1.8} \\
2455967 &  &  &    \\\hline
\end{tabular}
\\
$^{\ast}$for a description see Table~\ref{CCD_Kameras}.
\end{table}

\section{Summary and Discussion}

CVSO\,30 is a unique planetary system. For the first time both a close-in transiting and a wide directly imaged planet candidates are found to orbit a common host star. Furthermore, with $\sim$2.4\,Myrs it is among the youngest exoplanet systems. \\ CVSO\,30\,b is, for several reasons, a very interesting object. If confirmed as a planet it would be the first transiting planet orbiting a T-Tauri star and the first transiting planet found in a young open cluster. It also shows a highly variable shape and even the disappearance and reappearance of the event that could be due to orbital precession.\\ Our YETI monitoring observations of the young open cluster 25\,Ori started in 2010 January with the aim to detect and characterise young transiting planets. The YETI campaigns for 25\,Ori continued over three consecutive years until 2013 February, followed by two more years of photometric follow-up observations. Over this period of five years we obtained a significant collection of photometric time series of CVSO\,30. In total, our observations cover 62 time windows for the fading event with at least one data point. Out of these, 42 yielded usable data. However, one LC (JD 2455627) shows no indication for a brightness dip at the calculated time although the whole window was covered with observations. Our data show the previously claimed changing depth, disappearance and reappearance of the fading event as expected for a precessing planetary orbit. The values for the possible precession period range from $\sim$\,200 to 800\,d \citep{2013ApJ...774...53B,2015arXiv150604829K}. However, given the time span between a clear detection and a non-detection of the fading event the period could well be even much shorter. Our search for a periodic change in inclination yielded $P_{\mathrm{Incl}}$\,=\,152.97\,$\pm$\,0.55\,epochs ($\sim$\,68.5\,d) with a FAP of 0.2 per cent. However, considering the quality of the data and the gaps between seasons this period might be an alias of the observing cadence. But since \citet{2012ApJ...755...42V} calculated a possible precession period on the order of tens to hundreds of days our result is not inconsistent. Recently, \citet{2016MNRAS.457.3769H} repeated the precession modelling using an improved version of the Barnes model. Since they solution  requires stellar properties that disagree with the observations they claimed that precession might not be the right explanation. However, all the precession modelling by \citet{2013ApJ...774...53B}, \citet{2015arXiv150604829K} and \citet{2016MNRAS.457.3769H} is based on one phase-folded and binned LC per season (2009,2010). Since the shape of the LCs seem to change on faster time-scales than one year, averaging with subsequent fitting down to strong details might not be the best approach, in particular in the presence of star spots.\\ \citet{2015arXiv150608719C} report on a non-detection of the ading event at JD 2456284. Our observation in 2013 January (JD 2456305) and the one of \citet{2015arXiv150902176Y} from 2012 December 15 yielded a clear detection. These fading events are only $\sim$\,8\,d earlier and $\sim$\,21 later than the non-detection of \citet{2015arXiv150608719C}. The detection of \citet{2015arXiv150608719C} in 2013 November is in agreement with our photometric follow-up observations four epochs ($\sim$\,1.8\,d) later. The existence of the brightness dip in season 2013/2014 is also confirmed by \citet{2015arXiv150902176Y}. In the 2014/2015 season \citet{2015MNRAS.450.3991K} and \citet{2015arXiv150604829K} reported on four and five non-detections in 2015 January, respectively. However, the times of their observing window was based on the \citet{2012ApJ...755...42V} ephemeris. Considering our updated ephemeris (Eq. \ref{Elemente_CVSO30}) and the fact that the brightness dip happened even earlier in the 2014/2015 season, the fading event was not covered in their observations. In the case of \citet{2015MNRAS.450.3991K} signs of a brightness dip might be visible around 1\,h before their predicted time (see their Fig. 2 LC JD 2457026, JD 2457028, JD 2457029).\\ Orbital precession could also explain the very shallow fading event and the non-detection of the secondary eclipse in the observations with Spitzer \citep{2015arXiv150608719C}. The upper planetary radius limit deduced from the Spitzer data seem to be in disagreement with the transit modelling of the optical data. However, if the orbital inclination at this epoch is low enough that only part of planetary disc transits the star, the brightness dip will be very shallow. \\ Assuming the orbital precession, we modelled all detected fading events with the inclination and the mid-time as free parameters. Our derived stellar and planetary parameters are smaller but within the error bars in agreement with the values in \citet{2012ApJ...755...42V}.\\ In our period search we found no significant periodic signal in the O--C diagram. \\ The analysis of the out-of-event measurements yielded a stellar rotation period of $\sim$\,0.5\,d or a multiple of that. This is in agreement with the finding of \citet{2015MNRAS.450.3991K}. However, we cannot find the second fundamental period reported by \citet{2015MNRAS.450.3991K}. Hence, we cannot confirm that the period of the fading event is locked to the stellar rotation. Interestingly, the amplitude of light variation caused by stellar rotation seem to be correlated with the depth of the fading event. Although the depth changes for every observation, it seem to be generally higher in seasons with higher stellar activity (see Tables~\ref{CVSO30_results} and \ref{rotation_period}). Because of the low number statistics this cannot be quantified with the available data set.\\ A crucial point in the discussion is the confirmation of the planetary nature of CVSO\,30\,b. Because of the missing secondary transit and Rossiter-McLaughlin effect, the wavelength-dependent  depth, and the failing precession models, a planet with 3-5$M_{Jup}$ is disfavoured by \citet{2015arXiv150902176Y} and \citet{2016MNRAS.457.3769H}. Although a shorter precession period might be possible the correlation between depth and amplitude of variability weakens the giant-planet hypothesis.\\ \citet{2012ApJ...755...42V}, \citet{2015arXiv150608719C} as well as \citet{2015arXiv150902176Y} intensively discussed the possibility of stellar spots that mimic a transit signal. \citet{2015arXiv150608719C}  concluded that it is not impossible for an active low-mass star but it is very unlikely that a spot appears, evolves and disappears at approximately the same longitude stable over several years.\\ \citet{2015AJ....149..130S} found periodic brightness dips in the LCs of several classical T-Tauri stars that originate from the co-rotation of clumps of dust in the circumstellar disc. In these cases the stellar rotation period was close to the period of the brightness dip. Although \citet{2007ApJ...667..308C} found that $\sim$\,20 per cent of the WTTSs (in their sample) show evidence for the presence of a circumstellar disc, \citet{2007ApJ...671.1784H} did not detect any infrared excess in the Spitzer IRAC bands ($3.6, 4.5, 5.8$ and $8\mu$m) and, hence, considered CVSO\,30 as discless. This finding was confirmed by \citet{2015arXiv150902176Y} who reanalysed the SED of the star. Therefore dust clumps orbiting this star seem to be unlikely. However, \citet{2015arXiv150608719C} and \citet{2015arXiv150902176Y} showed that there are evidences for ongoing accretion (accretion feature in the optical light curve, strength and breadth of the H$\alpha$ line profile). This points to the existence of an accretion disc. Therefore the brightness dip could be caused by irregularities in the accretion disc, extinction by infalling material from the inner edge of the disc, or by a hotspot that is produced by accretion. The latter scenario is favoured by \citet{2015arXiv150902176Y} because it is consistent with a low accretion rate from a optical thin accretion disc.\\ An alternative explanation may be an occultation of the star by dust from a disintegrating planet. So far there are $\sim$\,100 planet candidates known with orbital periods shorter than one day, most of them smaller than twice the size of Earth. \citet{2013ApJ...779..165J} hypothesised that the origin of these very short period Earth like planets may be the remnants of disrupted hot jupiters. Two examples of disintegrating planets have been studied by \citet{2012ApJ...752....1R} and \citet{2015arXiv150404379S}. As in the case of CVSO\,30, they found a highly variable transit depth ranging from 0 to 1.3 per cent in a highly erratic manner. In both cases the actual planet is much smaller than expected from the transit depth. If the mass and radius of CVSO\,30\,b are grossly overestimated it may explain the missing signal in the RV and Rossiter-McLaughlin measurements of \citet{2015arXiv150608719C} and \citet{2015arXiv150902176Y}. The latter did not detect the spectroscopic transit in their observations from 2013 December 12, although a transit event is seen in our data one epoch before and after their observations (see Fig.~\ref{LC_CVSO30_Follow_up} transits JD 2456638 and 2456639). Furthermore, disintegrating planets show variations in the transit shape and asymmetric profiles, which can also be seen in some of our LCs (e.g. JD 2455890 or 2456611). \citet{2015arXiv150404379S} showed that the transit depth of disintegrating planets is significantly wavelength dependent which we cannot confirm with our observations (see Fig.~\ref{2456988}) but is seen in four out of five multicolour LCs of \citet{2015arXiv150902176Y}. The depth is decreasing with wavelength as expected for extinction by small dust grains. This is consistent with the shallower fading event in the IR observed by Spitzer. In seasons with higher stellar activity more material is 'blown' away from the planet through higher levels of stellar high-energy irradiation. Hence, a larger portion of the star could be occulted by dust resulting in a deeper brightness dip. The disintegrating planet hypothesis was also disfavoured by \citet{2015arXiv150902176Y} because of the fast orbital decay they detected. However, we cannot confirm this fast change in period of the fading event. CVSO\,30 seem to share some features with the object reported by \citet{2015Natur.526..546V}. They found six statistically significant groups of fading events in the light curve of the white dwarf WD\,1145+017. The O--C diagram reported by \citet{2015arXiv151006434C} shows the different groups of transits that are similar in period but phase shifted. \citet{2015Natur.526..546V} concluded that the white dwarf is likely transited by multiple disintegrating minor planets or planetesimals. The O--C diagram of CVSO\,30 might be explained by a similar mechanism, and hence, a disintegrating planet remains a possibility to explain the CVSO\,30 system.\\ In conclusion, the system is too complex to confirm the planetary nature of CVSO\,30\,b, yet. If it is indeed a giant planet on a precessing orbit the period may be shorter than previously thought. Since this explanation seem to be unlikely our most favoured solution would be a disintegrating planet (or planetesimals) or ongoing episodic accretion. We are continuing the process of analyzing the full data set obtained with all 13 telescopes during the three years of YETI monitoring. Additionally, we are also obtaining  further follow-up observations of this unique and fascinating system. CVSO\,30 would benefit greatly from continuous space-based observations like K2 \citep{2014AAS...22322801H}, or ESA's space telescope PLATO \citep['PLAnetary Transits and Oscillations of stars',][]{2011JPhCS.271a2084C}. If it is indeed confirmed as a planet, it will provide important constraints on planet formation and migration time-scales, and their relation to protoplanetary disc lifetimes.


\section*{Acknowledgements}

We would like to thank H. Gilbert, N. Tetzlaff, A. Berndt, T. Eisenbeiss, T. Roell, F. Giessler, S. Minardi, S. Fiedler, W. Bykowski, I. H\"{a}usler, P. N\"{a}hrlich, N. Chakrova, E. Schmidt, C. Ebbecke, Keppler, Menzel, Belko, and Fruhnert for participating in some of the observations at the University Observatory Jena and O. Contreras, D. Cardozo, F. Moreno, G. Rojas, R. Rojas, U. Sanchez, L. Araque, D. Gonz\'{a}lez and J. Moreno, at CIDA.\\ SR is currently a Research Fellow at ESA/ESTEC. SR, CA, RE, MK and RN would like to thank DFG for support in the Priority Programme SPP 1385 on the ``First Ten Million Years of the Solar system'' in projects NE 515/34-1 and -2, NE 515/33-1 and -2, and NE 515/35-1 and -2. TK acknowledges support by the DFG program CZ 222/1-1 and RTG 1351 (Extrasolar planets and their host stars). MK would like to thank Ronald Redmer and DFG in project RE 882/12-2 for financial support. MF acknowledges financial support from grants AYA2014-54348-C3-1-R and AYA2011-30147-C03-01 of the Spanish Ministry of Economy and Competivity (MINECO), co-funded with EU FEDER funds. DK and VR acknowledge support by project RD 08-81 of Shumen University. Wu, Z.Y. was supported by the Chinese National Natural Science Foundation grant No. 11373033. This work was also supported by the joint fund of Astronomy of the National Nature Science Foundation of China and the Chinese Academy of Science, under Grant U1231113. Zhou, X. was supported by the Chinese National Natural Science Foundation grands No. 11073032, and by the National Basic Research Program of China (973 Program), No. 2014CB845704,and 2013CB834902. MM and CG acknowledge DFG for support in program MU2695/13-1. JS, RN and MMH would like to thank the DFG for support from the SFB-TR 7. CG, and TOBS would like to thank DFG for support in project NE 515/30-1. CM acknowledges support from the DFG through grant SCHR665/7-1. RN would like to thank the German National Science Foundation (Deutsche Forschungsgemeinschaft, DFG) for general support in various projects. We would like to acknowledge financial support from the Thuringian government (B 515-07010) for the STK CCD camera used in this project. This work has been supported by a VEGA Grant 2/0143/13 of the Slovak Academy of Sciences. \\ This work is based on observations obtained with telescopes of the University Observatory Jena, operated by the Astrophysical Institute of the Friedrich-Schiller-University Jena.\\ The observations obtained with the MPG 2.2m telescope were supported by the Ministry of Education, Youth and Sports project - LG14013 (Tycho Brahe: Supporting Ground-based Astronomical Observations). We would like to thank the observers S.Ehlerova and A.Kawka for obtaining the data.\\ Based on observations obtained at the Llano del Hato National Astronomical Observatory of Venezuela, operated by CIDA for the Ministerio del Poder Popular para la Ciencia y Tecnolog{\'\i}a.

\bibliographystyle{mn2e}
\bibliography{literatur}


\appendix
\onecolumn

\section{LCs}

\begin{figure*}
 \centering
  \includegraphics[width=0.21\textheight, angle=270]{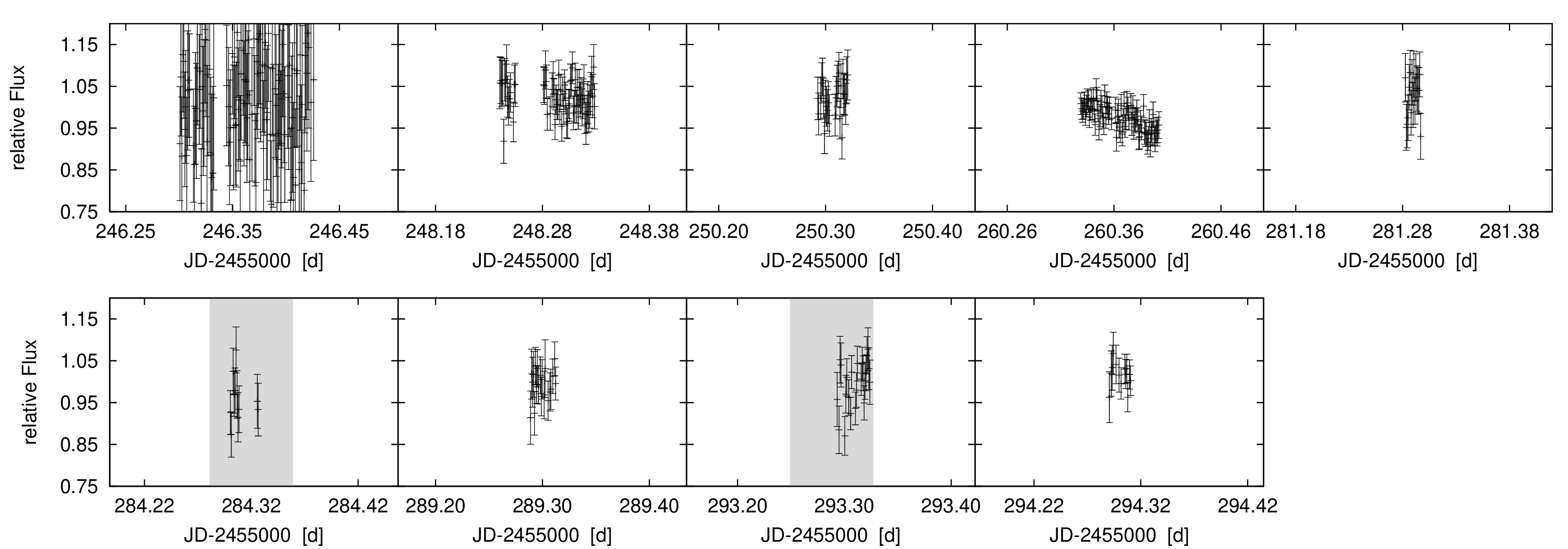}
  \caption{All LCs of CVSO\,30 observed in $R$ band in 2010 Jan-Apr (Season 0) with the 90\,cm telescope at the Unversity Observatory Jena. The grey shaded areas indicate the time window of the fading event fixed at measured period and mid-tim at epoch zero (see section \ref{Transit_timing}). The duration was fixed to the value in \citet{2012ApJ...755...42V}. The first panel is shown as an example for non-optimal observing conditions. The uncertainties will preempt every detection. This night was initially thrown out by the preceding LC treatment.}
\label{LC_S00}
\end{figure*}

\begin{figure*}
 \centering
  \includegraphics[width=0.95\textheight, angle=270]{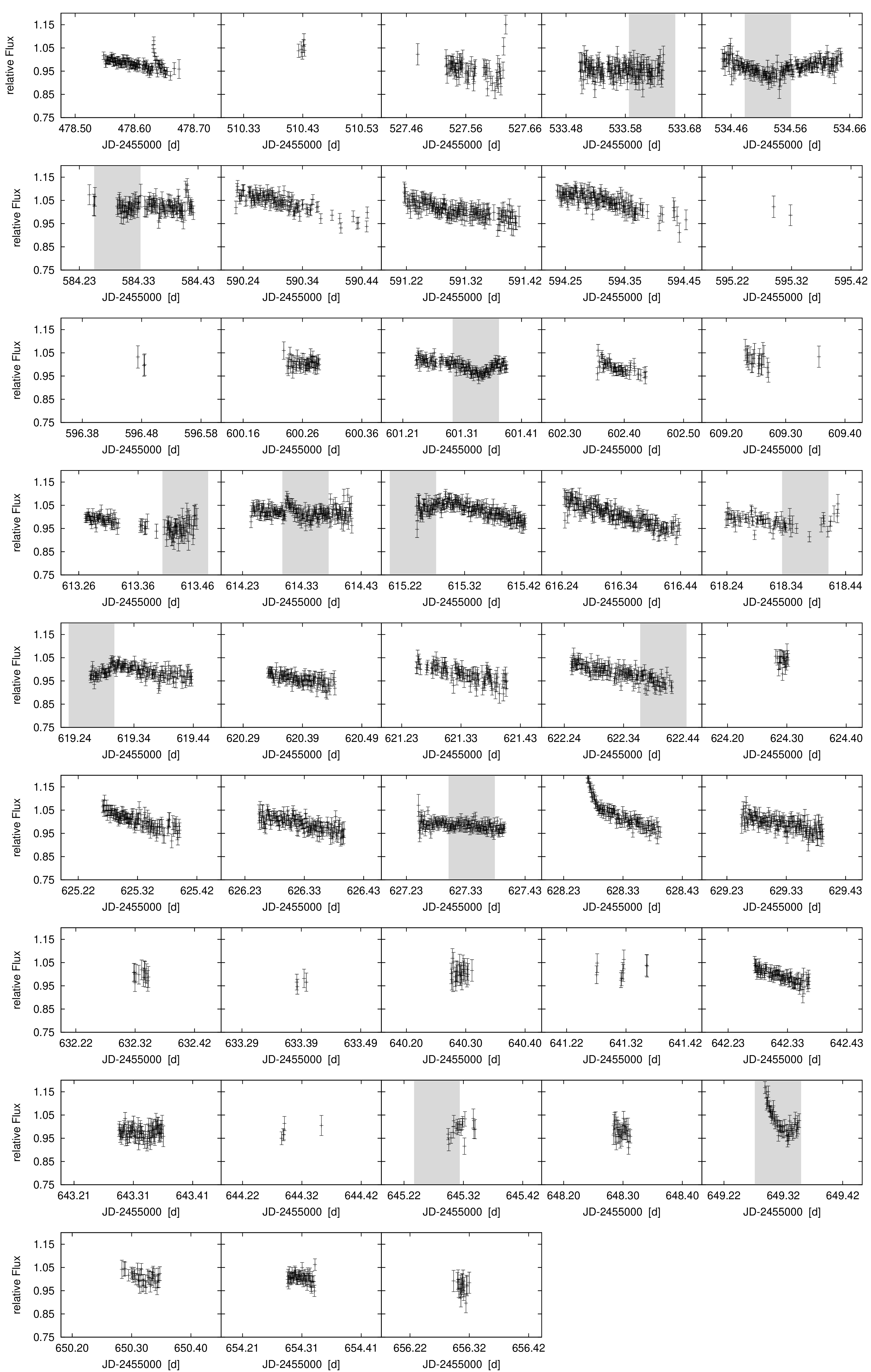}
  \caption{Same as Figure \ref{LC_S00} but for all LCs from Season 1 (2010/2011 season).}
\label{LC_S01_50s}
\end{figure*}

\begin{figure*}
 \centering
  \includegraphics[width=0.84\textheight, angle=270]{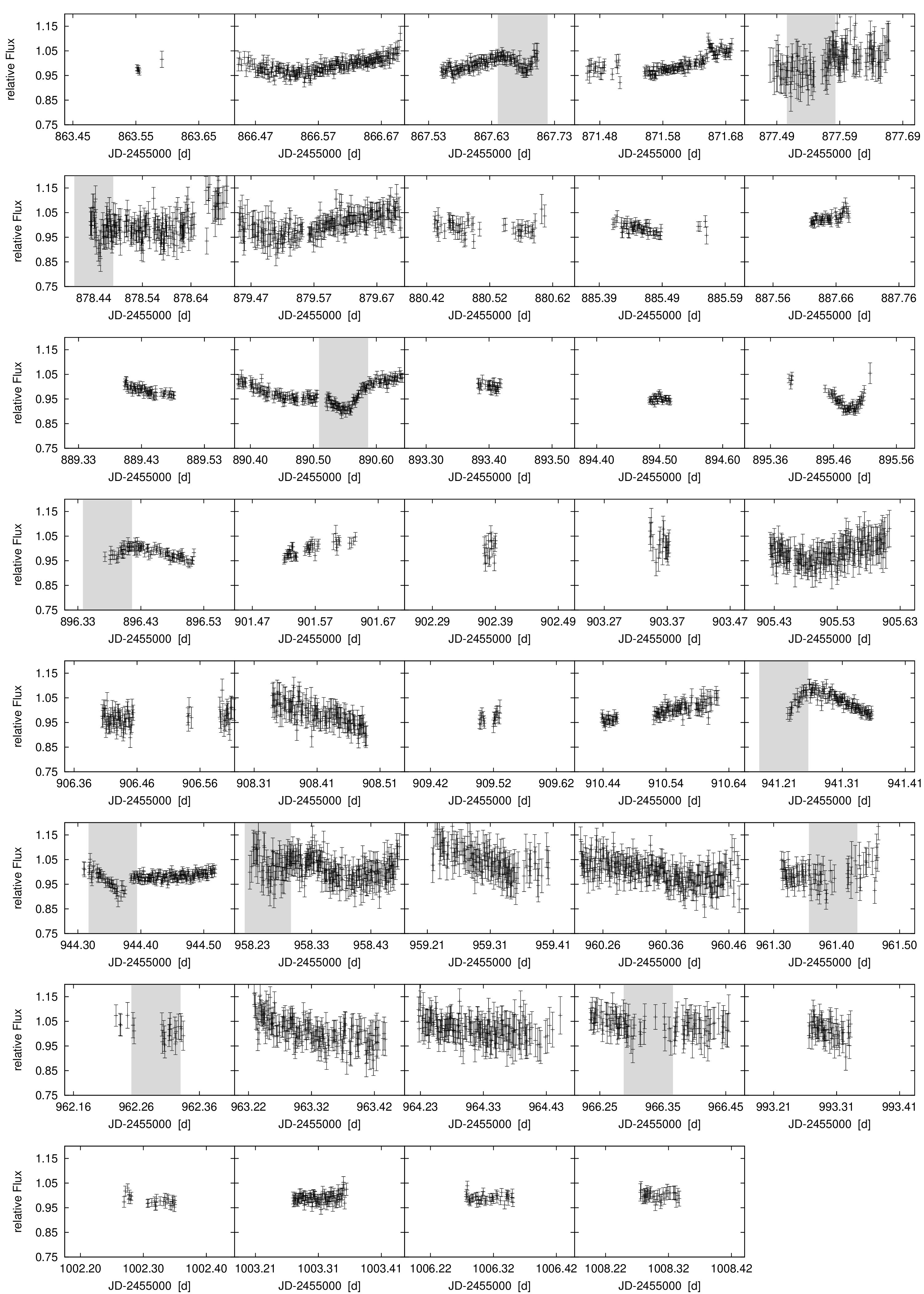}
  \caption{Same as Figure \ref{LC_S00} but for all LCs from Season 2 (2011/2012 season).}
\label{LC_S02_GSH}
\end{figure*}

\begin{figure*}
 \centering
  \includegraphics[width=0.106\textheight, angle=270]{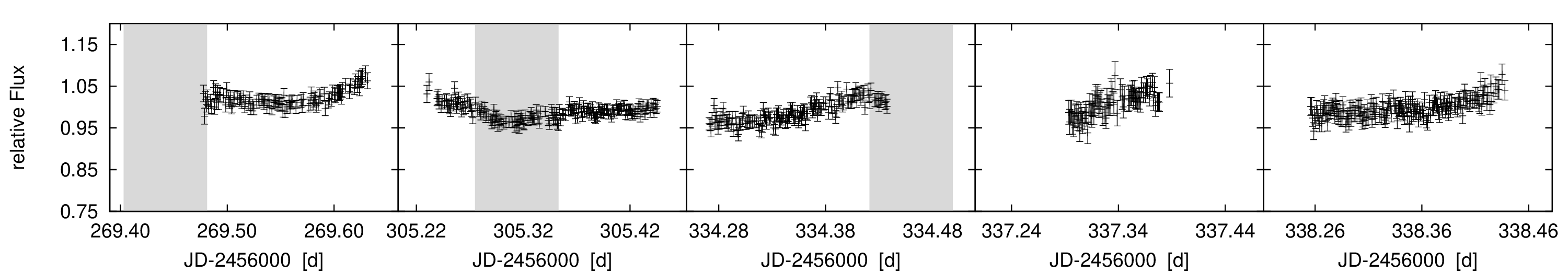}
  \caption{Same as Figure \ref{LC_S00} but for all LCs from Season 3 (2012/2013 season).}
\label{LC_S03_GSH}

 \centering
  \includegraphics[width=0.106\textheight, angle=270]{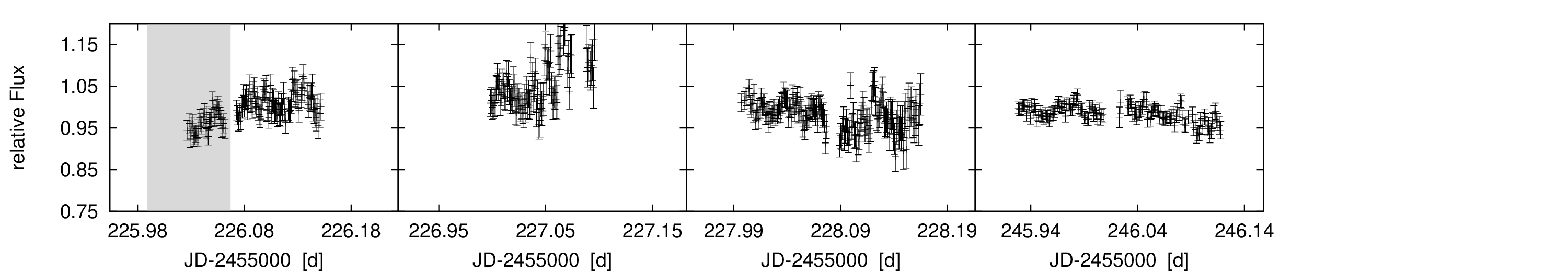}
  \caption{Same as Figure \ref{LC_S00} but for all LCs from Season 0 Gunma}
\label{LC_S00_Gunma}

 \centering
  \includegraphics[width=0.424\textheight, angle=270]{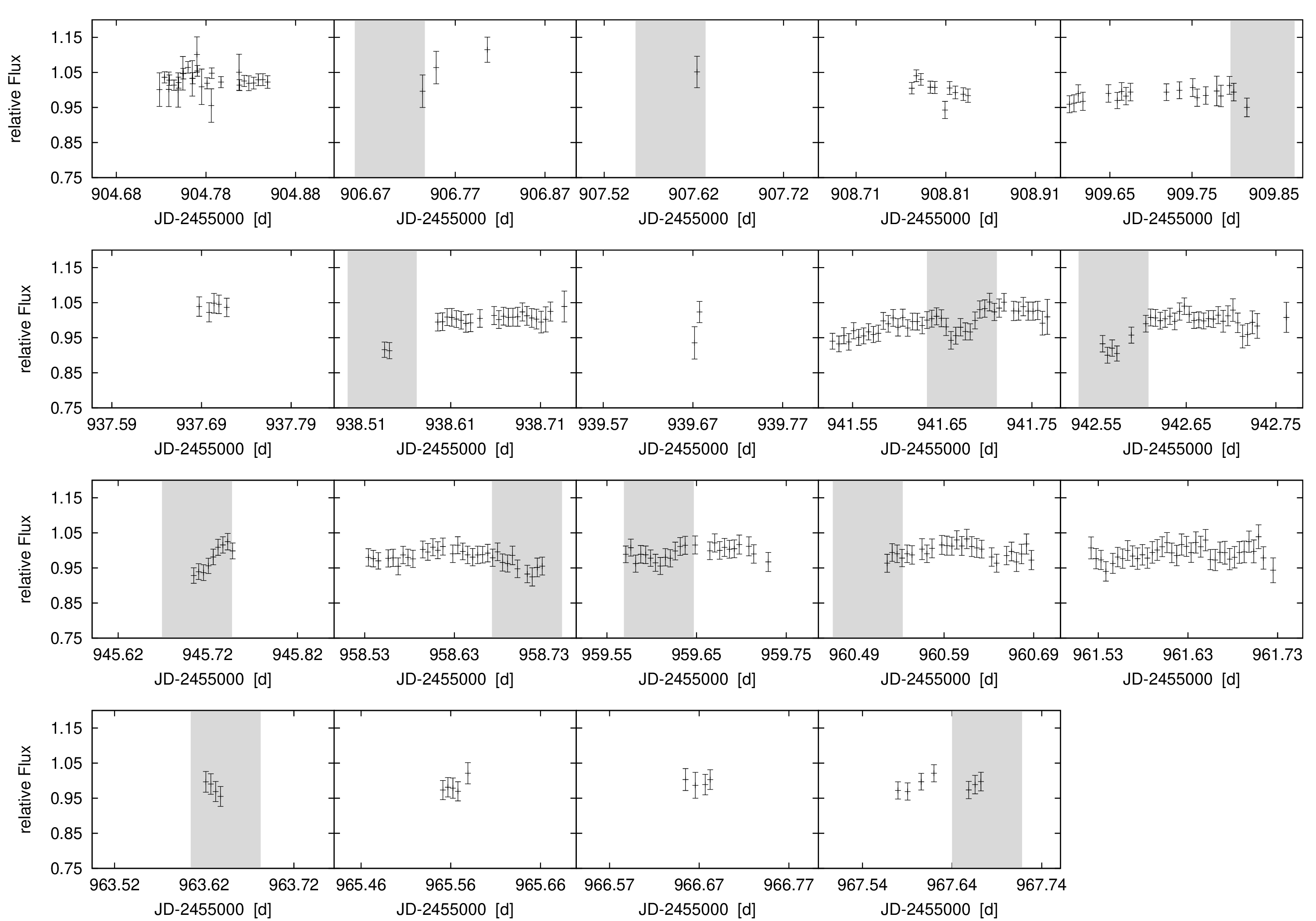}
  \caption{Same as Figure \ref{LC_S00} but for all LCs from Season 2 CIDA }
\label{LC_S02_CIDA}

 \centering
  \includegraphics[width=0.212\textheight, angle=270]{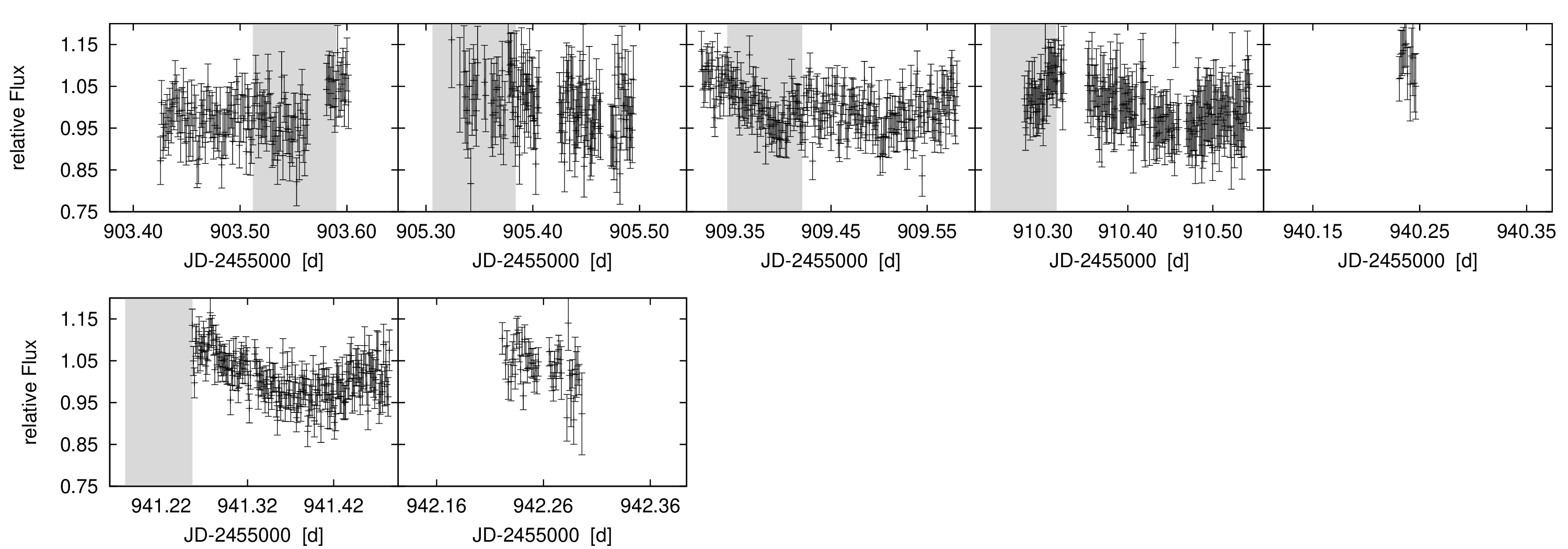}
  \caption{Same as Figure \ref{LC_S00} but for all LCs from Season 2 Rozhen }
\label{LC_S02_Rozhen}
\end{figure*}

\begin{figure*}
 \centering
  \includegraphics[width=0.212\textheight, angle=270]{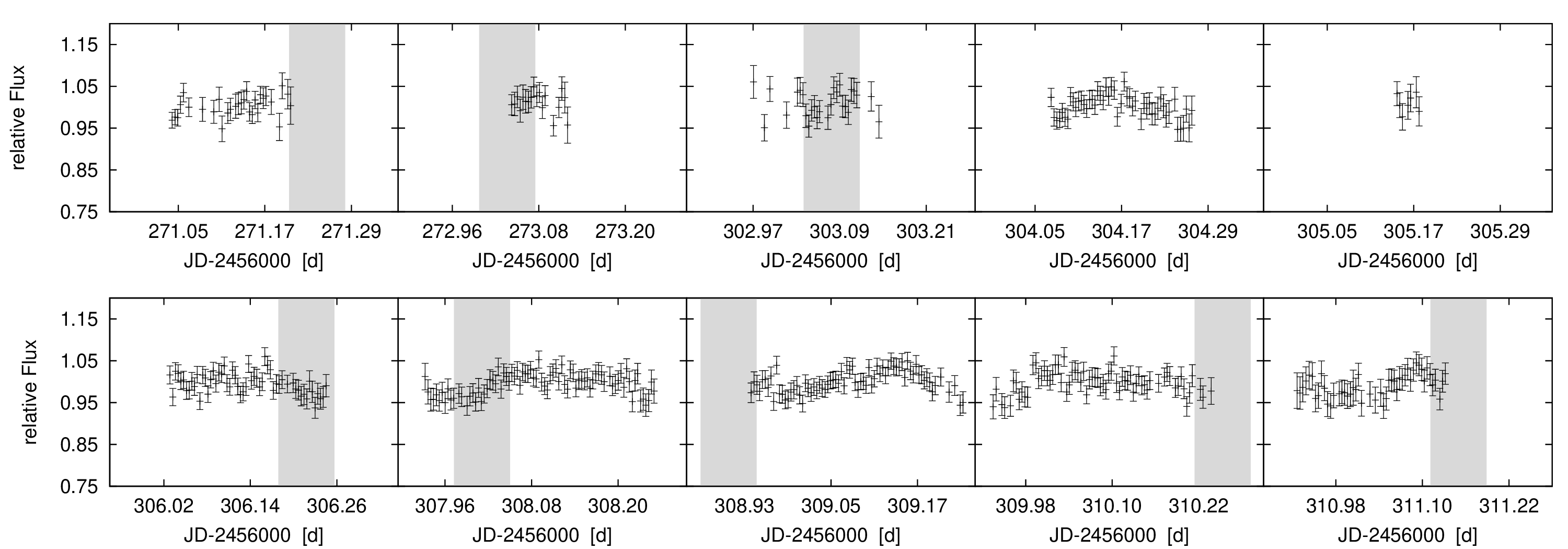}
  \caption{Same as Figure \ref{LC_S00} but for all LCs from Season 3 Xinglong}
\label{LC_S03_Xinglong}

  \includegraphics[width=0.212\textheight, angle=270]{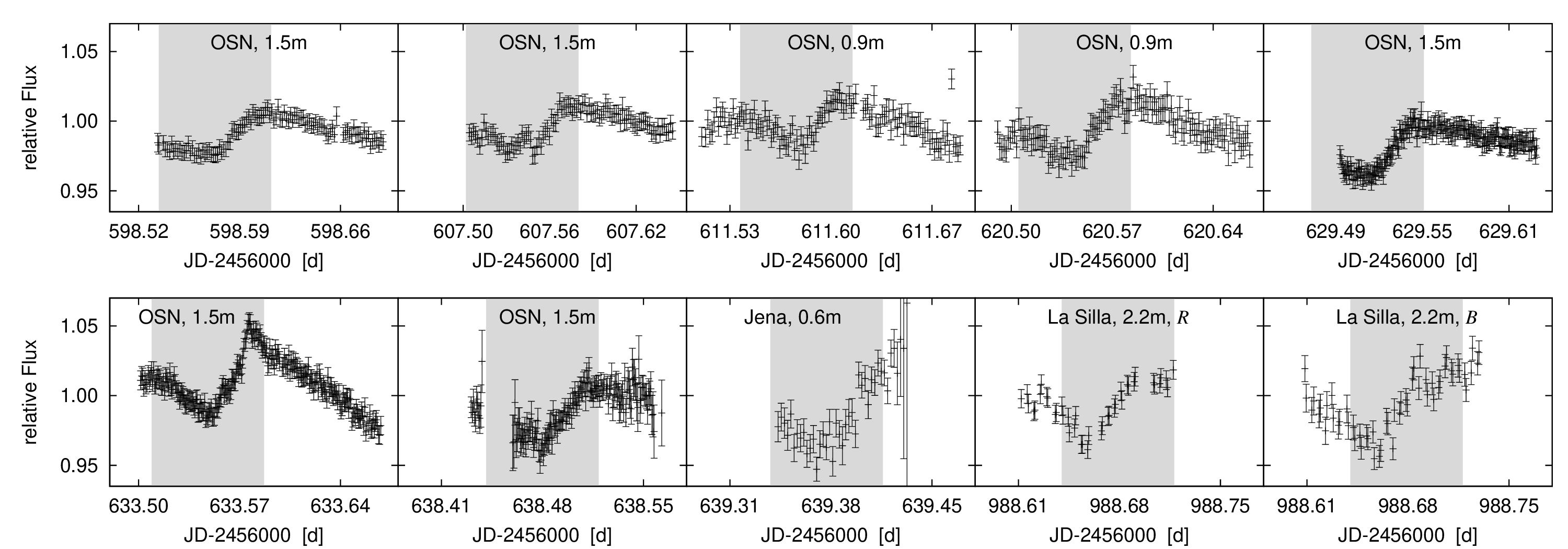}
  \caption{Same as Figure \ref{LC_S00} but for all follow-up LCs from the 2013/2014 Season. The used telescope are indicated. Unless noted otherwise, all LC are taken in $R$-band.}
\label{LC_CVSO30_Follow_up}

\end{figure*}

\section{Result of the transit fitting}

\begin{table}
\caption{Result of the transit fitting for all 38 detected fading events. The O--C was calculated with the ephemeris given in equation (\ref{Elemente_CVSO30}). The c and p in the second column mark the complete and partial fading events, respectively. The root-mean-square ($rms$) of the fit and the S/N of the observation are also given.}
\label{CVSO30_results}
\begin{tabular}{ccccccccc}
\hline \hline
Epoch &  & $T_{\mathrm{c}}$ (BJD$_{\mathrm{TDB}}$) & Inclination ($^{\circ}$) & Impact parameter & Depth (\%) & O--C (min) & $rms$ & S/N \\ \hline
-709	& p & 2455226.0343$^{+0.0110}_{-0.0120}$ & 77.6$^{+8.7}_{-10.0}$ &	0.39$^{+0.08}_{-0.09}$ & 4.7$^{+0.4}_{-0.5}$ & 8.5$^{+15.8}_{-17.3}$ 		& 26.62	& 1.8	\\
-23	& p & 2455533.6385$^{+0.0064}_{-0.0064}$ & 71.7$^{+7.5}_{-5.9}$ &	0.57$^{+0.07}_{-0.05}$ & 4.4$^{+1.0}_{-1.0}$ & 13.8$^{+9.2}_{-9.2}$ 		& 32.82	& 1.5	\\
-21	& c & 2455534.5281$^{+0.0038}_{-0.0032}$ & 64.6$^{+2.0}_{-1.3}$ &	0.77$^{+0.02}_{-0.01}$ & 3.7$^{+0.2}_{-0.2}$ & 3.5$^{+5.5}_{-4.6}$ 		& 23.01	& 1.6	\\
90	& p & 2455584.2957$^{+0.0130}_{-0.0140}$ & 63.5$^{+1.9}_{-5.7}$ &	0.81$^{+0.02}_{-0.05}$ & 3.5$^{+1.0}_{-1.0}$ & -3.0$^{+18.7}_{-20.2}$ 		& 25.19	& 2.2	\\
128	& c & 2455601.3394$^{+0.0016}_{-0.0020}$ & 63.4$^{+0.9}_{-1.2}$ &	0.81$^{+0.01}_{-0.01}$ & 3.5$^{+0.2}_{-0.2}$ & 3.6$^{+2.3}_{-2.9}$ 		& 16.19	& 2.2	\\
155	& p & 2455613.4205$^{+0.0260}_{-0.0110}$ & 64.6$^{+13.0}_{-4.5}$ &	0.77$^{+0.11}_{-0.04}$ & 3.7$^{+1.0}_{-1.0}$ & -33.3$^{+37.4}_{-15.8}$ 		& 22.74	& 1.2	\\
157	& p & 2455614.3294$^{+0.0130}_{-0.0046}$ & 64.6$^{+5.1}_{-4.6}$ &	0.77$^{+0.04}_{-0.04}$ & 3.7$^{+1.0}_{-1.0}$ & -15.9$^{+18.7}_{-6.6}$ 		& 23.75	& 1.5	\\
159	& p & 2455615.2512$^{+0.0095}_{-0.0140}$ & 59.2$^{+2.4}_{-2.1}$ &	0.92$^{+0.02}_{-0.02}$ & 2.5$^{+1.0}_{-1.0}$ & 20.2$^{+13.7}_{-20.2}$ 		& 19.38	& 1.3	\\
168	& p & 2455619.2631$^{+0.0068}_{-0.0034}$ & 72.5$^{+8.7}_{-7.5}$ &	0.54$^{+0.08}_{-0.07}$ & 4.4$^{+1.0}_{-1.0}$ & -14.0$^{+9.8}_{-4.9}$ 		& 21.15	& 2.0	\\
175	& p & 2455622.4127$^{+0.0180}_{-0.0100}$ & 62.2$^{+10.0}_{-1.8}$ &	0.84$^{+0.09}_{-0.02}$ & 3.3$^{+1.0}_{-1.0}$ & 1.6$^{+25.9}_{-14.4}$ 		& 22.64	& 1.7	\\
722	& p & 2455867.6878$^{+0.0041}_{-0.0044}$ & 63.1$^{+1.8}_{-1.6}$ &	0.82$^{+0.02}_{-0.01}$ & 3.4$^{+1.0}_{-1.0}$ & 4.1$^{+5.9}_{-6.3}$ 		& 19.44	& 1.9	\\
744	& c & 2455877.5413$^{+0.0100}_{-0.0075}$ & 68.9$^{+16.0}_{-7.0}$ &	0.65$^{+0.14}_{-0.06}$ & 4.2$^{+0.8}_{-0.4}$ & -12.0$^{+14.4}_{-10.8}$ 		& 46.67	& 0.9	\\
746	& p & 2455878.4553$^{+0.0120}_{-0.0110}$ & 60.6$^{+3.7}_{-1.9}$ &	0.89$^{+0.03}_{-0.02}$ & 2.9$^{+1.0}_{-1.0}$ & 12.8$^{+17.3}_{-15.8}$ 		& 46.22	& 0.7	\\
773	& c & 2455890.5504$^{+0.0051}_{-0.0044}$ & 72.5$^{+9.5}_{-4.5}$ &	0.54$^{+0.09}_{-0.04}$ & 4.4$^{+0.5}_{-0.4}$ & -4.0$^{+7.3}_{-6.3}$ 		& 16.82	& 3.6	\\
786	& p & 2455896.3727$^{+0.0052}_{-0.0041}$ & 78.1$^{+8.0}_{-7.4}$ &	0.37$^{+0.08}_{-0.07}$ & 4.7$^{+1.0}_{-1.0}$ & -13.8$^{+7.5}_{-5.9}$ 		& 13.86	& 3.7	\\
802	& c & 2455903.5600$^{+0.0110}_{-0.0091}$ & 70.5$^{+11.0}_{-3.9}$ &	0.60$^{+0.10}_{-0.04}$ & 4.3$^{+0.5}_{-0.4}$ & 4.8$^{+15.8}_{-13.1}$ 		& 53.95	& 0.9	\\
806	& p & 2455905.3500$^{+0.0130}_{-0.0160}$ & 76.1$^{+10.0}_{-7.6}$ &	0.43$^{+0.09}_{-0.07}$ & 4.6$^{+1.0}_{-1.0}$ & -0.3$^{+18.7}_{-23.0}$ 		& 73.17	& 0.7	\\
815	& c & 2455909.3959$^{+0.0074}_{-0.0040}$ & 65.7$^{+5.7}_{-2.5}$ &	0.74$^{+0.05}_{-0.02}$ & 3.8$^{+0.4}_{-0.3}$ & 14.6$^{+10.7}_{-5.8}$ 		& 47.34	& 0.9	\\
817	& p & 2455910.2808$^{+0.0074}_{-0.0058}$ & 81.7$^{+5.3}_{-5.9}$ &	0.26$^{+0.05}_{-0.06}$ & 4.8$^{+1.0}_{-1.0}$ & -2.6$^{+10.7}_{-8.4}$ 		& 48.85	& 1.2	\\
886	& p & 2455941.2124$^{+0.0077}_{-0.0048}$ & 74.3$^{+8.1}_{-5.6}$ &	0.49$^{+0.08}_{-0.05}$ & 4.5$^{+1.0}_{-1.0}$ & -13.9$^{+11.1}_{-6.9}$ 		& 15.76	& 3.2	\\
887	& c & 2455941.6711$^{+0.0058}_{-0.0084}$ & 64.6$^{+5.5}_{-1.7}$ &	0.77$^{+0.05}_{-0.01}$ & 3.7$^{+0.4}_{-0.2}$ & 0.9$^{+8.4}_{-12.1}$ 		& 16.82	& 2.9	\\
889	& p & 2455942.5768$^{+0.0073}_{-0.0150}$ & 77.8$^{+8.9}_{-5.8}$ &	0.38$^{+0.08}_{-0.05}$ & 4.7$^{+1.0}_{-1.0}$ & 13.8$^{+10.5}_{-21.6}$ 		& 21.55	& 2.7	\\
893	& c & 2455944.3656$^{+0.0018}_{-0.0028}$ & 66.7$^{+1.3}_{-1.7}$ &	0.71$^{+0.01}_{-0.02}$ & 3.9$^{+0.2}_{-0.3}$ & 6.9$^{+2.6}_{-4.0}$ 		& 17.32	& 2.8	\\
924	& c & 2455958.2821$^{+0.0054}_{-0.0063}$ & 79.8$^{+4.8}_{-9.2}$ &	0.32$^{+0.05}_{-0.09}$ & 4.7$^{+0.4}_{-0.5}$ & 30.2$^{+7.8}_{-9.1}$ 		& 44.71	& 1.2	\\
925	& p & 2455958.7195$^{+0.0085}_{-0.0076}$ & 67.0$^{+16.0}_{-4.5}$ &	0.71$^{+0.14}_{-0.04}$ & 4.0$^{+1.0}_{-1.0}$ & 14.3$^{+12.2}_{-10.9}$ 		& 12.42	& 3.8	\\
927	& p & 2455959.5966$^{+0.0057}_{-0.0058}$ & 78.3$^{+5.8}_{-7.9}$ &	0.37$^{+0.05}_{-0.07}$ & 4.7$^{+1.0}_{-1.0}$ & -14.0$^{+8.2}_{-8.4}$ 		& 12.65	& 4.2	\\
1698	& c & 2456305.3111$^{+0.0034}_{-0.0034}$ & 61.4$^{+1.3}_{-0.8}$ &	0.86$^{+0.01}_{-0.01}$ & 3.1$^{+0.1}_{-0.2}$ & -13.7$^{+4.9}_{-4.9}$ 		& 11.28	& 2.8	\\
1700	& p & 2456306.2220$^{+0.0068}_{-0.0056}$ & 79.0$^{+7.4}_{-8.1}$ &	0.34$^{+0.07}_{-0.08}$ & 4.7$^{+1.0}_{-1.0}$ & 6.6$^{+9.8}_{-8.1}$ 		& 23.40	& 2.1	\\
1704	& c & 2456308.0085$^{+0.0081}_{-0.0083}$ & 58.5$^{+1.6}_{-1.6}$ &	0.94$^{+0.01}_{-0.01}$ & 2.2$^{+0.2}_{-0.2}$ & -3.6$^{+11.7}_{-12.0}$ 		& 21.15	& 1.1	\\
2352	& c & 2456598.5721$^{+0.0020}_{-0.0017}$ & 59.1$^{+0.4}_{-0.7}$ &	0.93$^{+0.00}_{-0.01}$ & 2.4$^{+0.1}_{-0.1}$ & -0.6$^{+2.9}_{-2.4}$ 		& 3.55	& 7.7	\\
2372	& c & 2456607.5400$^{+0.0022}_{-0.0020}$ & 64.3$^{+1.8}_{-2.1}$ &	0.78$^{+0.02}_{-0.02}$ & 3.6$^{+0.2}_{-0.2}$ & -0.7$^{+3.2}_{-2.9}$ 		& 5.69	& 13.4	\\
2381	& c & 2456611.5790$^{+0.0019}_{-0.0024}$ & 62.0$^{+1.7}_{-0.9}$ &	0.85$^{+0.01}_{-0.01}$ & 3.2$^{+0.2}_{-0.2}$ & 4.2$^{+2.7}_{-3.5}$ 		& 6.67	& 5.1	\\
2401	& c & 2456620.5471$^{+0.0017}_{-0.0017}$ & 63.5$^{+0.9}_{-1.2}$ &	0.80$^{+0.01}_{-0.01}$ & 3.5$^{+0.2}_{-0.2}$ & 4.5$^{+2.4}_{-2.4}$ 		& 6.69	& 5.5	\\
2421	& p & 2456629.5150$^{+0.0014}_{-0.0018}$ & 61.0$^{+0.5}_{-0.4}$ &	0.88$^{+0.00}_{-0.00}$ & 3.0$^{+0.1}_{-0.1}$ & 4.4$^{+2.0}_{-2.6}$ 		& 4.86	& 8.2	\\
2430	& c & 2456633.5502$^{+0.0027}_{-0.0018}$ & 63.0$^{+0.8}_{-1.2}$ &	0.82$^{+0.01}_{-0.01}$ & 3.4$^{+0.1}_{-0.2}$ & 4.0$^{+3.9}_{-2.6}$ 		& 9.06	& 6.4	\\
2441	& p & 2456638.4813$^{+0.0015}_{-0.0017}$ & 63.5$^{+0.8}_{-0.7}$ &	0.80$^{+0.01}_{-0.01}$ & 3.5$^{+0.2}_{-0.2}$ & 2.0$^{+2.2}_{-2.4}$ 		& 8.95	& 3.9	\\
2443	& c & 2456639.3832$^{+0.0046}_{-0.0028}$ & 63.4$^{+2.1}_{-1.0}$ &	0.81$^{+0.02}_{-0.01}$ & 3.5$^{+0.2}_{-0.2}$ & 9.4$^{+6.6}_{-4.0}$ 		& 11.72	& 1.9	\\
3222$^{a}$	& c & 2456988.6606$^{+0.0034}_{-0.0034}$ & 62.2$^{+2.9}_{-2.5}$ &	0.84$^{+0.02}_{-0.02}$ & 3.3$^{+0.2}_{-0.2}$ & -25.3$^{+4.9}_{-4.9}$ 	& 9.25	& 2.8	\\
3222$^{b}$	& c & 2456988.6606$^{+0.0034}_{-0.0034}$ & 61.9$^{+2.4}_{-2.7}$ &	0.85$^{+0.02}_{-0.02}$ & 3.2$^{+0.2}_{-0.2}$ & -25.3$^{+4.9}_{-4.9}$ 	& 10.29	& 2.8	\\
\hline \hline
\end{tabular}
\\
$^{a}$ $R$-band \\
$^{b}$ $B$-band
\end{table}

\end{document}